\documentclass[%
 aip,
amsmath,amssymb,
reprint,%
]{revtex4-1}

\usepackage{graphicx}
\usepackage{dcolumn}
\usepackage{bm}
\usepackage{latexsym,bm}
\usepackage{amsmath,amsfonts}
\usepackage{amssymb}
\usepackage{mathtools}
\usepackage{mathrsfs}

\usepackage[utf8]{inputenc}
\usepackage[T1]{fontenc}
\usepackage{mathptmx}

\usepackage{float}
\usepackage{mathrsfs}
\usepackage[lofdepth,lotdepth]{subfig}
\usepackage{color}
\renewcommand{\vec}[1]{\mathbf{#1}}

\begin{document}

\preprint{AIP/123-QED}
\title{A Kinetic Equation for Particle Transport in Turbulent Flows}

\author{De-yu ZHONG}
\thanks{Corresponding author: zhongdy@tsinghua.edu.cn} 
\affiliation{State Key Laboratory of Hydroscience and Engineering 
	\\Tsinghua University, Beijing 100084, China }
\affiliation{Joint-Sponsored State Key Laboratory of Plateau Ecology and Agriculture\\ Qinghai University, Xining 810016, China}
\author{Guang-qian WANG}
\affiliation{Joint-Sponsored State Key Laboratory of Plateau Ecology and Agriculture\\ Qinghai University, Xining 810016, China}
\affiliation{State Key Laboratory of Hydroscience and Engineering 
	\\Tsinghua University, Beijing 100084, China }
\author{Tie-jian LI}
\affiliation{State Key Laboratory of Hydroscience and Engineering 
	\\Tsinghua University, Beijing 100084, China }

\author{Ming-xi ZHANG}
\affiliation{State Key Laboratory of Hydroscience and Engineering 
	\\Tsinghua University, Beijing 100084, China }
\author{You XIA}
\affiliation{State Key Laboratory of Hydroscience and Engineering 
	\\Tsinghua University, Beijing 100084, China }

\date{\today}

\begin{abstract}
One key issue in the probability density function (PDF) approach for disperse two-phase turbulent flows is to close the diffusion  term in the phase space.  This study aimed to derive a kinetic equation for particle dispersion in turbulent flows by ensemble averaging over all possible realisations of state transition paths in the phase space. The probability density function is expanded as a series in terms of the cumulants of particle paths in the phase space, by introducing a local path density operator to identify the distribution of particle paths. The expansion enables us to directly obtain a kinetic equation with the diffusion term in closed form. The kinetic equation derived in this study has following features that: (1) it has its coefficients expressed as functions of the cumulants of particle paths in the phase space;  (2) it applies to particle dispersion by non-Gaussian random forcing with long correlation time scales; (3) it presents new mechanisms responsible for particle diffusion. An application of the kinetic equation is also presented in this paper.

\end{abstract}

\maketitle

\section{\label{sec:level1}Introduction}



Flows of  continuous fluids carrying dispersed solid grains, gas bubbles, or liquid droplets are generally  referred to as disperse two-phase flows.  
Disperse two-phase flows occur in a rich variety of circumstances, including suspension of sediment particles in natural rivers and channels, dust storms in the atmosphere, mixing of bubbles or droplets in the chemical engineering devices, among many others. As an important discipline of fluid mechanics, disperse two-phase flows have been extensively  studies in the past decades.

Conventionally, there are two major categories of approaches have been developed to formulate disperse two-phase flows. The first category includes the methods referred to as the two-fluid models\cite{drew1999theory}, of which particle clouds are idealized as continuum media as their carrier fluids and the governing equations for both phase are derive based on the fundamental conservation laws for mass, momentum, and energy. 
Two-fluid models have been extensively investigated since the late 1970's, and successful applications tackling problems associated with dispersive particles in turbulent flows have been widely reported \cite{hsu2003two,hsu2004two,zhong2011transport,zhongdrif,ZHONG2015285}. 

However, as a direct descendant of continuum theory, two-fluid models for disperse two-phase flows suffer difficulties in closing the governing equations with well-founded constitutive relations \cite{drew1999theory}.  This situation is largely due to the challenge in incorporating the microscopic dynamics of particles into the macroscopic governing equations for particle-laden flows \cite{prosperetti1996disperse,zhang1997momentum}.  
This challenge was partly solved by another important category of  approaches based on stochastic theory or kinetic therry\cite{prosperetti1996disperse,zhang1997momentum,reeks1980eulerian,reeks1983transport,Reeks1991Kinetic,reeks1992continuume,Pandya2003Noniso,reeks2005on,DEREVICH1990631,derevich1994statistical,SWAILES199738,Hyland1999PDF,zaichik1999statistical,derevich2000statistical,derevich2001influence,zaichik2004probability,derevich2006statistical, zaichik2010modelling, zaichik2011statistical,MINIER20011,bragg2012particle,Pozorski1999PDFmodeling,Pandya2003Noniso}, of which the particle kinetic equation, or the particle probability density function (PDF) equation, is derived and applied to derive the macroscopic conservation equations for solid phase with constitutive relations properly defined. In the PDF formulation of disperse two-phase flows, a key issue is to close a correlation term arising from ensemble average on the equation for conservation of fine-grained probability density in the phase space. The well-known approaches include  LHDI theory \cite{reeks1980eulerian,reeks1983transport,Reeks1991Kinetic,reeks1992continuume,Pandya2003Noniso,reeks2005on}, functional method \cite{DEREVICH1990631,derevich1994statistical,SWAILES199738,Hyland1999PDF,zaichik1999statistical,derevich2000statistical,derevich2001influence,zaichik2004probability,derevich2006statistical, zaichik2010modelling, zaichik2011statistical,MINIER20011,bragg2012particle}, and cumulant expansion method \cite{Pozorski1999PDFmodeling,Pandya2003Noniso}.

The principal difficulty encountered in closing the turbulent correlation term is to formulate diffusion of particles as a result of their random motion in turbulent flows. Because turbulence of fluids often leads to long-time correlation in flow fields\cite{Reeks1991Kinetic, bragg2012particle}, particle motion cannot always be regarded as a Markovian process driven by white noise; instead, in many cases, they are dispersed by random forcing to exhibit a strong memory effect of non-Markovian dynamics\cite{Reeks1991Kinetic}.  In order to take into account non-Markovianity of particle motion, for instance, in the LHDI approximation, the RGT invariance is imposed on particle motion along its path in the phase space to remove the restriction on correlation time scales\cite{reeks1980eulerian,reeks1983transport,Reeks1991Kinetic,reeks1992continuume,Pandya2003Noniso,reeks2005on}; while in the functional methods, the diffusion coefficients are expressed in terms of the functional derivative of particle path with respect to random impulse in the phase space to allow for influences of long-time correlation on particle diffusion\cite{DEREVICH1990631,derevich1994statistical,SWAILES199738,Hyland1999PDF,zaichik1999statistical,derevich2000statistical,derevich2001influence,zaichik2004probability,derevich2006statistical, zaichik2010modelling, zaichik2011statistical,MINIER20011,bragg2012particle}.  Applications show that the kinetic equation methods are successful in modelling disperse two-phase flows, although there is much room still remained for further investigation in deriving a closed kinetic equation for disperse two-phase flows. Moreover, the previously reported studies also provided us a fundamental concept that, in deriving a Eulerian probability distribution function for particle motion in turbulent flows, combination with a Lagrangian point of view is necessary to consider memory effect of non-Markovian dynamics in turbulent diffusion of particles.

This paper aims to derive a kinetic equation for particle diffusion in turbulent flows with a new method to formulate particle diffusion in the phase space. For particles dispersed by random forcing with a long correlation time scale, considering the fact that it is the state transition paths (or the trajectories of particles in the phase space) along which memory effect takes place to influence future particle states before losing coherence, one infers that statistical characteristics of particle motion are determined not only by the system states, but also by the state transition paths and their distribution. In this paper, a local path density operator is introduced to identify state transition paths. The local path density operator is also a fine-grained probability density function in the phase space, but it differs from previous studies in that, it is extended to serve as a indicator of both state and state transition path in fine-grained scale. In the derivation, the probability density function is given by an ensemble average of the local path density operator, and by expanding it as a series in terms of the cumulants of particle paths in the phase space, a kinetic equation with the diffusion term in closed form is obtained directly. It shows that the derived kinetic equation possesses the ability to account for non-Markovianity in particle motion driven by non-Gaussian random forces; while in the white noise limit, it is contracted to the classical Fokker-Planck equation. Moreover, it leads to finding two new mechanisms responsible for diffusion in the phase space. 

In the following sections, we firstly provide a detailed derivation of a kinetic equation for a general stochastic system; thereafter, two important properties of the kinetic equation, specifically, the memory effect and the Markovian approximation, are discussed, followed by an application of this study on dispersion of particles in homogeneous turbulent flows. Concluding remarks are presented in the final section.


%
%
\section{\label{sec:level2}Formulation}

%
%
\subsection{\label{sec:level2-1} Local path density operator}

Consider a system described by its state variable $\vec{X}=\{{X}_j\}$ with $N$ components, in which each component $X_j$ ($j=1,\cdots, N$) can be either a vector or a scalar, depending on the specific problem of interest. For instance, if one-point PDF model is considered, $X_j$ ($j=1,\cdots, 2$) denotes position and velocity vector of a particle, respectively. The following derivation can be easily extended to a many-particle PDF model.  In this paper, a state transition path is elaborated mathematically as a curve in the phase space, along which the system changes its state  from $\vec{X}_s$ at time $s$ to arrive at $\vec{X}$ at time $t$, and is denoted by $\vec{X}(t)=\vec{X}(t|\vec{X}_s,s)$. Furthermore, we assume that $\vec{X}(t)$ is differentiable, or at least piecewise differentiable, with respect to time $t$, and it observes that 
\begin{eqnarray}\label{eq-1}
\dot{\vec{X}}\equiv\frac{\mathrm{d}\vec{X}}{\mathrm{d} t} = H(\vec{X}), 
\end{eqnarray}
where $H$ is an arbitrary integrable function of the variable $\vec{X}$ and the first-order system of Eq. \eqref{eq-1} is assumed to have an initial state (value) of  
\begin{eqnarray}\label{eq-2}
\vec{X}_s   = \vec{X}(s|\vec{X}_s,s),
\end{eqnarray}
at time $s(<t)$. Since dilute disperse two-phase flows are considered, collisions between particles was not taken into account in this study. 

Consider a special case that the system is in state $\vec{x}=\{x_1,x_2,\cdots x_N\}$ at the time $t$, on condition that it is in the state of $\vec{y}=\{{y}_{1}, {y}_{2},\cdots,{y}_{N}\}$ at the time $s$; that is, $\vec{X}_s=\vec{y}$. In this case, the state transition path is $\vec{X}=\vec{X}(t|\vec{y},s)$, with  $\vec{X}(t|\vec{x},t)=\vec{x}$ and $\vec{X}(s|\vec{y},s)=\vec{y}$.  
To identify those paths arriving at $\vec{x}$ at time $t$, we introduce an operator $\chi$, which serves to describe the density of the state transition paths that pass $\vec{y}$ at time $s$ to arrive at $\vec{x}$ at the time $t$, given by Eq. \eqref{eq-1}. It is defined as a function of distance $|\vec{x}-\vec{X}(t|\vec{y},s)|$:
\begin{eqnarray}\label{eq-3}
\chi(t)  &=& \chi(|\vec{x}-\vec{X}(t|\vec{y},s)|) 
=\prod_{j=1}^N \chi(|x_j-X_j(t|\vec{y},s)|). 
\end{eqnarray}
The local path density operator $\chi$ is assumed to have a sharp value at $\vec{x}=\vec{X}(t|\vec{y},s)$, while it is zero elsewhere. The most simple selection of $\chi$ is the Dirac-$\delta$ function, namely $\chi(t)=\delta(|\vec{x}-\vec{X}(t|\vec{y},s)|)$. But for the sake of generality, it is maintained as a general functional of the state difference $|\vec{x}-\vec{X}(t|\vec{y},s)|$ as in Eq. {\eqref{eq-3}}.

From the definition of the local path density operator, we can infer that $ \chi$ depends on the path $\vec{X}=\vec{X}(t|\vec{y},s)$, which changes with time. The time rate of change of $\chi $ along the curve $\vec{X}(t|\vec{y},s)$ is
\begin{eqnarray}\label{eq-4}
\frac{\partial \chi(|\vec{x}-\vec{X}(t|\vec{y},s)|)}{\partial {t}} 
=
\dot{\vec{X}} \nabla_{\vec{X}} \chi(|\vec{x}-\vec{X}(t|\vec{y},s)|).
\end{eqnarray}
As $\chi=\chi(|\vec{x}-\vec{X}(t|\vec{y},s)|)$ is a function of $|\vec{x}-\vec{X}(t|\vec{y},s)|$, it is straightforward to verify the identity $\nabla_{\vec{X}} \chi=-\nabla_{\vec{x}}\chi$. By denoting $\mathscr{L}=\dot{\vec{X}}\nabla_{\vec{x}}=\sum_{j=1}^N{\dot{X}_j\partial/\partial x_j}$, Eq. \eqref{eq-4} can be written in an operator form as follows:  
\begin{eqnarray}\label{eq-5}
\frac{\partial \chi}{\partial {t}}=-\mathscr{L} \chi.
\end{eqnarray}

Eq. \eqref{eq-5} is a Liouville-type equation for the local path density operator $\chi$, which has an operator solution along the path $\vec{X}=\vec{X}(t|\vec{y},s)$ as \citep{van1992stochastic, zwanzig2001nonequilibrium, path-integrals}:
\begin{eqnarray}\label{eq-6}
\chi(|\vec{x}-\vec{X}(t|\vec{y},s)|)  
&=&\mathscr{U}(t|s) \chi(|\vec{x}-\vec{X}(s|\vec{y},s)|) \nonumber \\
&=&\mathscr{U}(t|s)\chi(|\vec{x}-\vec{y}|). 
\end{eqnarray}
Here, $\mathscr{U}(t|s)$ is a time evolution operator defined by \cite{van1992stochastic, path-integrals}:
\begin{eqnarray}\label{eq-7}
\mathscr{U}(t|s)
&=&\overleftarrow{T}\mathrm{e}^{-\int_s^{{t}} \text{d} \tau \mathscr{L}(\tau)} \nonumber \\
&=&\overleftarrow{T}\sum_{n=0}^{\infty}\frac{(-1)^n}{n!} \left(\int_{s}^{t}\text{d}\tau
\mathscr{L}(\tau) \right)^n,
\end{eqnarray} 
where $\overleftarrow{T}$ denotes the time-ordering operator by which the integrations in
\begin{eqnarray}\label{eq-8}
\left(\int_{s}^{t} \mathrm{d}\tau \mathscr{L} \right)^n  
=
\int_{s}^{t} \mathrm{d}\tau_1  \cdots\int_{s}^{t}\text{d}\tau_{n}
\mathscr{L}(\tau_1)  \cdots \mathscr{L}(\tau_{n}), 
\end{eqnarray}
for $n=1, 2, \cdots$, are correctly ordered so that the earlier times in the products of the integrand stand to the left of those with later times ($\tau_{1}>\tau_{2}>\cdots>\tau_{n}$).

\subsection{\label{sec:level2-2} Probability density function}
Generally, for a Markovian process, the possibility of finding a system having the state of $\vec{x}$ at the time $t$ is provided by the Chapman-Kolmogorov equation to maps the state $\vec{y}$ at the time $s$ to the state $\vec{x}$ at the time $t$ by means of the transition probability function $f(\vec{x},t|\vec{y},s)$ \citep{risken1984fokker}. In deriving the Fokker-Planck equation, the state transition probability $f(\vec{x},t|\vec{y},s)$ was formally expanded as a Taylor series of the transition moments; therefore, it has to be limited to cases in which the state transition time scale $|t-s|$ must be infinitesimal to make the mathematical definition of the state transition moments meaningful\citep{risken1984fokker}. However, when investigating a non-Markovian process, it must to be modified. 

In statistical mechanics, states of a system are regarded to  correspond to a set of different realisations that are compatible with their boundary constraints\cite{balescu,drew1999theory,zwanzig2001nonequilibrium}, which leads to any possible realisations of the system states, as well as their transition paths in the phase space, exhibiting a certain degree of uncertainties. The uncertainties, particularly those observed in the state transition paths, are the object of our focus in this study. 

According to the definition of the local path density operator, the ensemble average on it leads to $f(\vec{x},t|\vec{y},s)$, the conditional probability density function of finding the system in state $\vec{x}$ at the time $t$, given that it is in state $\vec{y}$ at the time $s$.
Let the state transition path given by $\vec{X}(\Gamma)=\vec{X}(t|\vec{y},s)$ corresponding to a realisation $\Gamma$, and  $F(\Gamma)$ being the distribution of the realisation $\Gamma$, which satisfies $\int \mathrm{d} \Gamma  F(\Gamma) =1$, the state transition probability density function $f(\vec{x},t|\vec{y},s)$, therefor, is given by
\begin{eqnarray}\label{eq-9}
f(\vec{x},t| \vec{y},s) 
&=&\int \mathrm{d} {\Gamma }F(\Gamma)
\chi(|\vec{x}-\vec{X}(\Gamma)|) \nonumber\\
&=&\langle \chi(|\vec{x}-\vec{X}(t|\vec{y},s)|)\rangle 
, 
\end{eqnarray}
where a variable closed by a pair of angles ``$\langle\rangle$" is its ensemble average, defined by: 
\begin{eqnarray}\label{eq-10}
\langle A \rangle  
&=&
\int \text{d}\Gamma F(\Gamma) A(\Gamma).
\end{eqnarray} 

Eq. \eqref{eq-9} indicates that the local path density operator $\chi$ assumes the function of picking up those paths leading from a given point $\vec{y}$ at the time $s$ to arrive at $\vec{x}$ at the time $t$ from all possible state transition paths. Since each realization $\Gamma$ corresponds to a path $\vec{X}(\Gamma)$, the ensemble average in Eq. \eqref{eq-9} is essentially taken on all of possible state transition paths leading from $\vec{y}$ to $\vec{x}$. This definition differs from classical statistical mechanics, in which a local density function is usually employed to identify points (states) of interest in the phase space, and thus the ensemble average in classical statistical mechanics is carried out on system states and their distribution\cite{balescu,zwanzig2001nonequilibrium}. This modification is significant, by which non-Markovianity in particle motion can be reflected in the state transition probability function given by Eq. \eqref{eq-9}. 

By substituting Eq. \eqref{eq-6} into Eq. \eqref{eq-9}, the conditional probability density function $f(\vec{x},t| \vec{y},s)$ can be written in an equivalent form as follows: 
\begin{eqnarray} \label{eq-11}
f(\vec{x},t| \vec{y},s)
&=& 
\int \text{d} \Gamma  F(\Gamma)\chi(|\vec{x}-\vec{X}(t|\vec{y},s)|) \nonumber \\
&=& \langle \mathscr{U}(t|s) \rangle \chi(|\vec{x}-\vec{y}|),  
\end{eqnarray}
where $\langle \mathscr{U}(t|s) \rangle$ is the path ensemble averaged time evolution operator, and is expanded in detail in the following manner:
\begin{eqnarray}\label{eq-12}
\langle\mathscr{U}(t|s) \rangle 
&=&
\int \text{d}\Gamma  F(\Gamma) \mathscr{U}(t|s) \nonumber \\
&=& 
\overleftarrow{T}\mathrm{exp}
\left(
\sum_{n=1}^{\infty}\frac{(-1)^n}{n!}
\left\langle\left\langle\left(
\int_s^{t}\text{d}\tau \mathscr{L} \right)^n\right\rangle\right\rangle\right),
\end{eqnarray}
in which $\langle\langle\rangle\rangle$ represents the cumulant operator; for example, $\langle\langle A \rangle\rangle= \langle A \rangle$ and  $\langle\langle AB \rangle\rangle= \langle(A-\langle A  \rangle )(B-\langle B  \rangle ) \rangle$ are the first- and second-order cumulant, respectively, regarding the ensemble average defined by Eq. \eqref{eq-10}. In the derivation of Eq. \eqref{eq-12}, we have used the result of the ensemble average of exponent function \citep{risken1984fokker,van1992stochastic}. 

Eq. \eqref{eq-11} is in essence a series expansion of the  state transition probability function $f(\vec{x},t| \vec{y},s)$ in terms of the cumulants regrading state transition paths, rather than state transition moments in the derivation of the classical Fokker-Planck equation \cite{risken1984fokker}. This difference is of significance, because it provides us with a new method to consider non-Markovian dynamics in particle diffusion. 

\subsection{\label{sec:level2-3} Path-averaged kinetic equation}
Differentiating both sides of Eq. \eqref{eq-11} with respect to $t$, and inserting $\mathscr{L}=\dot{\vec{X}}\nabla_{\vec{x}}$ into the resulting equation, we derive a new kinetic equation for $f(\vec{x},t|\vec{y},s)$ including an infinite number of terms as follows:
\begin{eqnarray} \label{eq-13}
\frac{\partial f(\vec{x},t|\vec{y},s)}{\partial {t}} 
&=& 
\sum_{n=1}^{\infty}(-1)^n\nabla_{\vec{x}}^n   \langle\mathscr{D}^{(n)}(\vec{x},t| \vec{y},s) \rangle f(\vec{x},t|\vec{y},s),\nonumber\\
\end{eqnarray}
where the coefficient $ \langle\mathscr{D}^{(n)}(\vec{x},t|\vec{y},s)\rangle $ is given by
\begin{eqnarray} \label{eq-14}
\langle \mathscr{D}^{(n)}(\vec{x},t|\vec{y},s)\rangle 
&=&
\frac{1}{n!}\frac{\partial}{\partial t}
\overleftarrow{T}
\left\langle\left\langle\left(
\int_s^{t}\text{d}\tau \dot{\vec{X}} \right)^n
\right\rangle\right\rangle.
\end{eqnarray}
The operator $\nabla_{\vec{x}}^n \langle \mathscr{D}^{(n)}(\vec{x},t|\vec{y},s)\rangle$ in Eq. \eqref{eq-13} is defined by    
\begin{eqnarray} \label{eq-15}
\nabla_{\vec{x}}^n \langle \mathscr{D}^{(n)}(\vec{x},t|\vec{y},s)\rangle 
&=&
\frac{\partial^n}{\partial x_{j_1}\cdots\partial x_{j_n}} \langle \mathscr{D}^{(n)}_{j_1\cdots j_n}(\vec{x},t|\vec{y},s)\rangle,
\end{eqnarray}
in which the summation convention with respect to the repeated subscript $j_\nu$ ($j_{\nu}=1, 2, \cdots,  N$, and $\nu=1, \cdots, n$) is used, and the partial differential operators apply to all of the following variables.  

$f(\vec{x},t| \vec{y},s)$ is a conditional distribution function. It is advisable to derive a probability density function $f(\vec{x},t)$ in numerous circumstances. In general, this can be obtained by the following relation: 
\begin{eqnarray} \label{eq-16}
f(\vec{x},t)=\int \mathrm{d}\vec{y}f(\vec{x},t| \vec{y},s)f(\vec{y},s).  
\end{eqnarray}  
However, Eq. \eqref{eq-16} leads to a conditional average on $\left\langle \mathscr{U}(t|s) \right\rangle $, making the resulting equation complicated. For this reason, it is useful to define the distribution function for $\vec{x}$ in a different manner. Multiplying both sides of Eq. \eqref{eq-9} with a Dirac $\delta$-function of $\delta(|\vec{y}-\vec{X}(s)|)$, where $\vec{X}(s)$ is a known value determined by tracking from the time $t$ through the path $\vec{X}(\tau)=\vec{X}(\tau|\vec{X}(s),s)$ ($s \le \tau \le t$) back to the time $s$ with the condition that $\vec{x}=\vec{X}(t|\vec{X}(s),s)$, i.e., $\vec{X}(s)=\vec{X}(s|\vec{x},t)$, using the relation Eq. \eqref{eq-11}, we find that 
\begin{eqnarray} \label{eq-17}
\int{\mathrm{d}\vec{y}}f(\vec{x},t|\vec{y},s)\delta(|\vec{y}-\vec{X}(s)|) 
=\langle\chi(|\vec{x}- {\vec{X}}(t)|)\rangle.
\end{eqnarray}

\begin{figure*}[htbp] 
	\centering
	\includegraphics[scale=0.45, trim=0cm 0.cm 0cm 0cm]{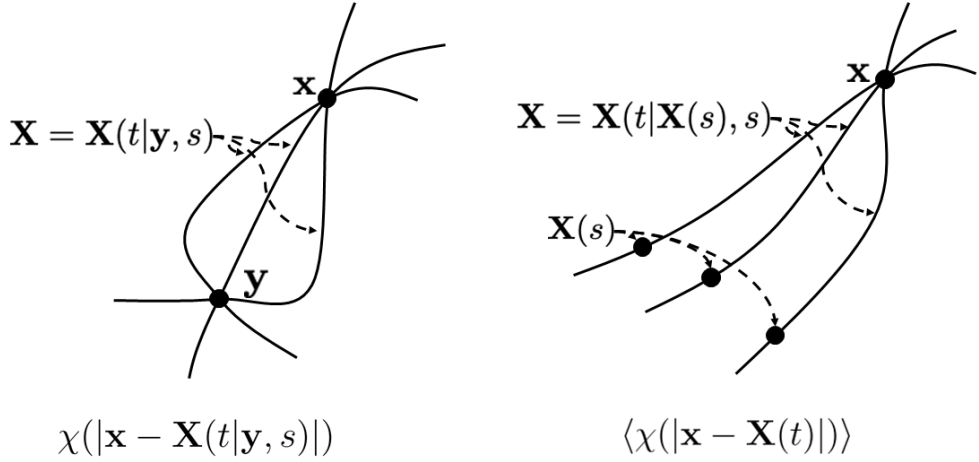} 
	\caption{\label{fig:Fig1}{{Schematic diagram of differences between $\langle \chi(|\vec{x}-\vec{X}(t|\vec{y},s)|)\rangle$ and $\langle\chi(|\vec{x}- {\vec{X}}(t) |)\rangle$. The figure illustrates that $\langle \chi(|\vec{x}-\vec{X}(t|\vec{y},s)|)\rangle$ is an ensemble path from a given state $\vec{y}$ at time $s$ to arrive at $\vec{x}$ at time $t$, while $\langle\chi(|\vec{x}- {\vec{X}}(t) |)\rangle$ is an ensemble taken on all paths leading to $\vec{x}$. Correspondingly, the time evolution operator $\langle \mathscr{U}(t|s)\rangle$ in Eq. \eqref{eq-11} involves the ensemble average on the integral curves starting from a given state $\vec{y}$, while in Eq. \eqref{eq-18}, the ensemble average is taken on the integral curves, with an initial state $\vec{X}(s)$.}}}
\end{figure*}

The major difference between $\langle \chi(|\vec{x}-\vec{X}(t|\vec{y},s)|)\rangle$ and $\langle\chi(|\vec{x}- {\vec{X}}(t) |)\rangle$ is depicted in Fig. \ref{fig:Fig1}. The figure illustrates that $\langle \chi(|\vec{x}-\vec{X}(t|\vec{y},s)|)\rangle$ is an ensemble average on the paths from a given state $\vec{y}$ at time $s$ to arrive at $\vec{x}$ at time $s$, while $\langle\chi(|\vec{x}- {\vec{X}}(t) |)\rangle$ is an ensemble average taken on all paths leading to $\vec{x}$. Moreover, $ \langle\chi(|\vec{x}- {\vec{X}}(t) |)\rangle$ is, in essence, the distribution function for $\vec{x}$ at time $t$; that is, 
\begin{eqnarray} \label{eq-18}
f(\vec{x},t) &\equiv& \langle \chi(|\vec{x}- {\vec{X}}(t)|)\rangle 
=\langle \mathscr{U}(t|s) \rangle \chi(|\vec{x}-\vec{X}(s)|).  
\end{eqnarray}  
It should be noted that the time evolution operator $\langle \mathscr{U}(t|s)\rangle$ in Eq. \eqref{eq-11} involves the ensemble average on the integral curves starting from a given state $\vec{y}$, while in Eq. \eqref{eq-18}, the ensemble average is taken on any curve leading to $\vec{x}$.

%
%

Using the result yielded by Eq. \eqref{eq-17}, multiplying both sides of Eq. \eqref{eq-13} with $\delta(|\vec{y}-\vec{X}(s)|)$, and integrating the resulting equation with respect to $\vec{y}$, we obtain a generalized kinetic equation for $f(\vec{x},t)$ as follows:  
\begin{eqnarray} \label{eq-19}
\frac{\partial f(\vec{x},t) }{\partial t} 
= 
\sum_{n=1}^{\infty}(-1)^n\nabla_{\vec{x}}^n  \langle \mathscr{D}^{(n)}(\vec{x},t) \rangle f(\vec{x},t),  
\end{eqnarray}
where the coefficient $\langle \mathscr{D}^{(n)}(\vec{x},t)\rangle$ involves the ensemble average on the integral curves starting from an initial state $\vec{X}(s)$, instead of a given state $\vec{y}$, as in Eq. \eqref{eq-13}. Eq. \eqref{eq-19} can also be derived directly by differentiating both sides of Eq. \eqref{eq-18} with respect to $t$.

Because it is derived by the ensemble average over all possible realisations of the particle paths in the phase space, Eqs. \eqref{eq-13} and \eqref{eq-19} can be referred to as path averaged kinetic equations.  It shows that Eq. \eqref{eq-13} (or Eq. \eqref{eq-19}) contains an infinite number of terms as the Kramers-Moyal expansion for Markovian processes\citep{risken1984fokker, Sto-m-cgispinGardiner}. According to the Pawula theorem, if the Kramers-Moyal expansion is not truncated at the second order, it must contain an infinite number of terms \citep{risken1984fokker, Sto-m-cgispinGardiner}. This is true for the current study, although the expansion coefficients are the functions of the cumulants \citep{Sto-m-cgispinGardiner}. 
When Eq. \eqref{eq-19} is truncated at $n=2$, we obtain a kinetic equation, as follows:
\begin{eqnarray} \label{eq-20}
\frac{\partial f(\vec{x},t) }{\partial t} 
+
\nabla_{\vec{x}} \langle \mathscr{D}^{(1)}  \rangle f(\vec{x},t) 
=
\nabla_{\vec{x}}^2 \langle \mathscr{D}^{(2)}  \rangle f(\vec{x},t). 
\end{eqnarray}

Eq. \eqref{eq-20} resembles the classical Fokker-Planck equation in form derived by truncating the Kramers-Moyal expansion at the second order. However, an essential difference exists. Eq. \eqref{eq-14} illustrates that $\langle\mathscr{D}^{(1)}\rangle$ is a path-averaged state transition velocity and $\langle\mathscr{D}^{(2)}\rangle$ is a measurement of dispersion of the state transition paths, rather than the jump moments in the Fokker-Planck equation\cite{risken1984fokker}. This difference is crucial which makes this study can be applied to diffusion due to non-Markovian motion of particles where memory effect is prominent. This point is further discussed in details in Section \ref{sec:level3}. It shows that Eq. \eqref{eq-20} can be alternatively written as an integro-differential equation with a memory kernel serving to allow for memory effect in non-Markovian processes.

%
%
\subsection{\label{sec:level2-4} Coefficients of kinetic equation}
The coefficient $\langle \mathscr{D}^{(n)}\rangle$ in Eq. \eqref{eq-19} can be expressed in a more compact and concise form. $\langle \mathscr{D}^{(n)}\rangle$ is a function of the integration of $\dot{\vec{X}}(t)$ along the curve $\vec{X}(t)=\vec{X}(t|\vec{X}(s),s)$ leading from $\vec{X}(s)$ to $\vec{x}$, denoted as  
\begin{eqnarray}\label{eq-21}
\vec{S}(t)  
=
\int_s^t \mathrm{d} \tau \dot{\vec{X}} = \int_C \mathrm{d} \vec{X}  
=
\left\{\int_C \mathrm{d} {X}_1,\cdots \int_C \mathrm{d} {X}_N  \right\}. 
\end{eqnarray}

On the one hand, the time ordering integration in Eq. \eqref{eq-17} can be expanded as follows \citep{van1992stochastic, zwanzig2001nonequilibrium, path-integrals}: 
\begin{widetext}
\begin{eqnarray}\label{eq-22}
\overleftarrow{T}  \left(\int_{s}^t \mathrm{d}\tau \dot{\vec{X}}(\tau)\right)^n
=
\int_{s}^t \mathrm{d}\tau_1 \cdots  \int_{s}^{\tau_{n-1}} \mathrm{d}\tau_n
\dot{\vec{X}}(\tau_1)\cdots\dot{\vec{X}}(\tau_n). 
\end{eqnarray}
\end{widetext}
On the other hand, denoting   
\begin{eqnarray}\label{eq-23}
{I}_n(t)= \int_{s}^t \mathrm{d}\tau_1 \cdots  \int_{s}^t \mathrm{d}\tau_n
\dot{\vec{X}}(\tau_1)\cdots\dot{\vec{X}}(\tau_n), 
\end{eqnarray}
and bearing in mind that the integrand $\dot{\vec{X}}(\tau_1)\cdots\dot{\vec{X}}(\tau_n)$ is symmetric in its arguments $\tau_1,\tau_2,\cdots,\tau_n$, it is straitforward that \cite{Joachain1975Quantum}
\begin{widetext}
\begin{eqnarray}\label{eq-24}
\int_{s}^t \mathrm{d}\tau_1 \cdots  \int_{s}^{\tau_{n-1}} \mathrm{d}\tau_n
\dot{\vec{X}}(\tau_1)\cdots\dot{\vec{X}}(\tau_n)=\frac{1}{n!}{I}_n(t)=\frac{1}{n!}\underbrace{\vec{S}(t)\cdots\vec{S}(t)}_n\equiv\frac{1}{n!}\vec{S}^n(t).
\end{eqnarray}
\end{widetext}
Therefore, $\left\langle \mathscr{D}^{(n)} (\vec{x},t) \right\rangle $ can be written in a simple form as follows:
\begin{eqnarray}\label{eq-25}
\langle \mathscr{D}^{(n)} (\vec{x},t)\rangle 
= \frac{1}{n! } \frac{\partial }{\partial t} \left\langle\left\langle  \frac{\vec{S}^n(t)}{n! } \right\rangle\right\rangle.
\end{eqnarray}

Eq. \eqref{eq-25} shows that $\langle \mathscr{D}^{(n)}\rangle$ is function of the cumulants with respect to the integral curves of the transition path $\vec{S}(t)$; expressing $\langle \mathscr{D}^{(n)}\rangle$ in terms of the correlation function $\left\langle\left\langle {\vec{S}^n } \right\rangle\right\rangle $ provides us with a new angle to view the manner of diffusion: it is the variation in the correlation of particle paths that drives a system state to diffuse in the phase space. Moreover, by means of the ensemble average on the state transition paths, the memory effct on $\vec{X}(t)$ of non-Markovianity can be considered, which is discussed in Section \ref{sec:level3-1}.  
Most importantly, since the coefficients of the kinetic equation derived in this paper are the functions of the cumulants of particle paths, which do not directly involve random forcing as a source term in the particle motion equations, the current study can be applied to stochastic systems where random forcing cannot be  separated linearly from deterministic accelerations.  This point is crucial when a system is driven by  non-linear stochastic accelerations.



%
%

\section{\label{sec:level3}Non-Markovianity and Markovian Approximation}

%
%
\subsection{\label{sec:level3-1}Non-Markovianity}

As has been mentioned in the introduction, accounting for non-Markovian behavior observed in dispersion of particles in turbulent flows lies at the heart of PDF approaches. In this paper, we developed a different approach which is expected to be able to consider memory effect of non-Markovian dynamics on particle diffusion. Therefore, a discussion about non-Markovianity presenting in the kinetic equation is crucial. 

To discuss the non-Markovianity taken into accounted by Eq. \eqref{eq-20}, an equivalent form of it is derived herein. We begin from Eq. \eqref{eq-16}. Differentiating it with respect to $t$, we obtained 
\begin{eqnarray}\label{eq-26}
\frac{\partial f(\vec{x},t)}{\partial t}
=
\int \mathrm{d}\vec{y} f(\vec{y},s) \frac{\partial f(\vec{x},t|\vec{y},s)}{\partial t}.
\end{eqnarray}
It can be demonstrated by means of the identity $\nabla_{\vec{x}}\chi(|\vec{x}-\vec{y}|)=-\nabla_{\vec{y}}\chi(|\vec{x}-\vec{y}|)$ that
\begin{eqnarray}\label{eq-27}
\frac{\partial f(\vec{x},t|\vec{y},s)}{\partial t}
&=&
\sum_{n=1}^{\infty}(-1)^n \nabla_{\vec{x}}^n  \left\langle \mathscr{D}^{(n)}(\vec{x},t|\vec{y},s)\right\rangle  f(\vec{x},t|\vec{y},s) \nonumber\\
&=&
\sum_{n=1}^{\infty} \nabla_{\vec{y}}^n  \left\langle \mathscr{D}^{(n)}(\vec{x},t|\vec{y},s)\right\rangle f(\vec{x},t|\vec{y},s). 
\end{eqnarray}

With the substitution of Eq. \eqref{eq-27} into \eqref{eq-26}, the drift term ($n=1$) is derived directly as
\begin{eqnarray}\label{eq-28}
\text{\emph{Drift}}
=-\nabla_{\vec{x}}\int \mathrm{d}\vec{y} \langle \mathscr{D}^{(1)}(\vec{x},t|\vec{y},s)\rangle f(\vec{x},t|\vec{y},s) f(\vec{y},s) 
,
\end{eqnarray}
and the diffusion term ($n=2$) is derived through integration by parts as
\begin{eqnarray}\label{eq-29}
\text{\emph{Diffusion}}
&=&
\nabla_{\vec{x}}\int \mathrm{d}\vec{y} \langle\mathscr{D}^{(2)}(\vec{x},t|\vec{y},s) \rangle f(\vec{x},t|\vec{y},s)\nabla_{\vec{y}}f(\vec{y},s)
.
\end{eqnarray}

In deriving Eqs. \eqref{eq-28} and \eqref{eq-29}, we use the following two identities: 
\begin{eqnarray}\label{eq-30}
\nabla_{\vec{x}}f(\vec{x},t|\vec{y},s)=-\nabla_{\vec{y}} f(\vec{x},t|\vec{y},s),
\end{eqnarray}
and 
\begin{eqnarray}\label{eq-31}
(-1)^{n-k} \nabla^{n-k}_{\vec{x}} \langle \mathscr{D}^{(n)}\rangle=\nabla^{n-k}_{\vec{y}}\langle \mathscr{D}^{(n)}\rangle,\quad k=1\cdots n.
\end{eqnarray}
Details in derivations of Eqs. \eqref{eq-30} and \eqref{eq-31} are given in Appendix \ref{sec:A-A} and \ref{sec:A-B}.  

Using Eqs. \eqref{eq-28} and \eqref{eq-29}, with the first- and second-order terms maintained in Eq. \eqref{eq-26}, we obtain the kinetic equation rewritten as an integro-differential equation (see details in Appendix \ref{sec:A-C}). It is : 
\begin{widetext}
\begin{eqnarray}\label{eq-32}
\frac{\partial f(\vec{x},t)}{\partial t}
&=&
-\nabla_{\vec{x}} \langle \dot{\vec{X}}(t)\rangle f(\vec{x},t) 
+\nabla_{\vec{x}}\int{\mathrm{d}\vec{y}}
\int_0^t \mathrm{d}s 
\mathscr{B}(\vec{x},t;\vec{y},s)
\nabla_{\vec{y}}f(\vec{y},s)
,
\end{eqnarray}
\end{widetext}
where 
\begin{eqnarray}\label{eq-33}
\langle \dot{\vec{X}}(t)\rangle=\frac{1}{f(\vec{x},t)} \int \mathrm{d} \vec{y} \langle \dot{\vec{X}}(t|\vec{y},s)\rangle f(\vec{x},t|\vec{y},s)f(\vec{y},s)
\end{eqnarray}
is the drift velocity, expressed as a conditional average of $\langle \dot{\vec{X}}(t|\vec{y},s)\rangle$, and 
\begin{eqnarray}\label{eq-34}
\mathscr{B}(\vec{x},t;\vec{y},s)=\frac{1}{2}f(\vec{x},t|\vec{y},s)\left\langle\left\langle \dot{\vec{X}}(s|\vec{y},s)\dot{\vec{X}}(t|\vec{y},s) 
\right\rangle\right \rangle
\end{eqnarray} 
is the memory kernel. 

Eq. \eqref{eq-32} is an integro-differential equation; although it is seldom applied in studying real problems in disperse two-phase flows, yet it is useful in discussion of non-Markovianity of a system. It is found that Eq. \eqref{eq-32} is similar in  form to that derived by \citet{zwanzig2001nonequilibrium} by the projection method, which has been demonstrated to be able to account for memory effect in non-Markovian processes by means of integration of the memory kernel, i.e., Eq. \eqref{eq-34}, over all previous states and time. Being an equivalent form to Eq. \eqref{eq-32}, therefore,  Eq. \eqref{eq-20} is able to be applied to circumstances where particle motion is non-Markovian.

%
%
\subsection{\label{sec:level3-2}Markovian approximation}
It is necessary to discuss if the present study can be reduced to classical Fokker-Planck equation when Markovian processes are considered. In that case, the correlation time $\Delta t= t-s$ is vanishingly small, i.e., the white noise limit, and thus the integral cure can be well approximated by a line segment as follows:
\begin{eqnarray}\label{eq-35}
\vec{S}(t) = \int_{s}^t \mathrm{d}\tau \dot{\vec{X}}  
 =  \Delta t \dot{\vec{X}}(t) +O(\Delta t) =\Delta \vec{X}(t) +O(\Delta t) .
\end{eqnarray}
At the same time, Eq. \eqref{eq-25} can be written equivalently as 
\begin{eqnarray}\label{eq-36}
\langle \mathscr{D}^{(n)} (\vec{x},t)\rangle
=\frac{1}{n!}\frac{1}{n!}\lim_{\Delta t \to 0}\frac{\left\langle\left\langle \vec{S}^n(t+\Delta t)\right\rangle\right\rangle-\left\langle\left\langle \vec{S}^n(t)\right\rangle\right\rangle}{\Delta t} 
.
\end{eqnarray}
Expanding $\vec{S}(t+\Delta t)$ at $t$ as a Taylor series, we obtained that, for $n=1$
\begin{eqnarray}\label{eq-37}
\langle \mathscr{D}^{(1)}(\vec{x},t) \rangle
&=&\lim_{\Delta t \to 0} \frac{\langle \vec{S}(t+\Delta t)\rangle-\langle\vec{S}(t)\rangle}{\Delta t}\nonumber\\
&=&\lim_{\Delta t \to 0} \frac{\langle \Delta \vec{X}\rangle}{\Delta t}=\langle\dot{\vec{X}}(t)\rangle.
\end{eqnarray}
Bearing in mind that $ \langle\Delta^2\vec{X}  \rangle\sim O(\Delta t)$ while $ \langle\Delta\vec{X}  \rangle^2 \sim O(\Delta^2 t)$ for Brownian particles\cite{risken1984fokker}, for $n=2$
\begin{eqnarray}\label{eq-38}
\langle \mathscr{D}^{(2)}(\vec{x},t) \rangle
&=&\frac{1}{2!}\frac{1}{2!}\lim_{\Delta t \to 0} \frac{\langle \langle\vec{S}^2(t+\Delta t)\rangle\rangle-\langle \langle\vec{S}^2(t)\rangle\rangle}{\Delta t} \nonumber \\
&=&\frac{1}{2} \lim_{\Delta t \to 0}\frac{  \langle\Delta^2\vec{X}  \rangle -\langle\Delta\vec{X}  \rangle^2 }{\Delta t}\nonumber \\
&=&\frac{1}{2} \lim_{\Delta t \to 0}\frac{  \langle\Delta^2\vec{X}  \rangle}{\Delta t}.
\end{eqnarray}

Eqs. \eqref{eq-37} and \eqref{eq-38} are the Markovian approximation of Eq. \eqref{eq-25} for $n=1$ and $n=2$ respectively. As a comparison, the coefficient of the Kramers-Moyal expansion for Markovian processes is given by\citep{risken1984fokker}:
\begin{eqnarray}\label{eq-39}
\langle \mathscr{D}^{(n)}\rangle  
&=& 
\frac{1}{n!}\lim_{\Delta t \to 0} \frac{\langle \Delta^n \vec{X} \rangle}{\Delta t}  \nonumber\\
&=& 
\frac{1}{n!}\lim_{\Delta t \to 0} \frac{1}{\Delta t} \langle (\vec{X}(t+\Delta t)-\vec{X}(t))^n \rangle,
\end{eqnarray} 
which means that, for $n=1$,  
\begin{equation}\label{eq-40}
\langle \mathscr{D}^{(1)} \rangle
=\lim_{\Delta t \to 0} \frac{ \langle \Delta \vec{X} \rangle}{\Delta t} 
=\lim_{\Delta t \to 0} \frac{1}{\Delta t} \langle(\vec{X}(t+\Delta t)-\vec{X}(t)) \rangle, 
\end{equation}
and for $n=2$,
\begin{equation}\label{eq-41}
\langle \mathscr{D}^{(2)} \rangle
=\lim_{\Delta t \to 0} \frac{ \langle \Delta^2 \vec{X} \rangle}{\Delta t}
=\frac{1}{2}\lim_{\Delta t \to 0} \frac{1}{\Delta t} \langle (\vec{X}(t+\Delta t)-\vec{X}(t)|)^2 \rangle
.
\end{equation}

When compared to  Eq. \eqref{eq-40} to \eqref{eq-41}, given the correlation time $t-s$ is infinitesimal to make $\vec{S} = \Delta \vec{X}$ for Markovian processes, it is found that Eq. \eqref{eq-37} and \eqref{eq-38} are the same as that of the classical Fokker-Planck equation. This result means that the Markovian approximation of the kinetic equation derived in this paper is the classical Fokker-Planck equation.  Actually, this conclusion can also be found by Markovian approximation of  Eq. \eqref{eq-32}. For Markovian processes, because $f(\vec{x},t|\vec{y},s)= \delta (\vec{x}-\vec{y})(1+O(\Delta t)) $ (see Eq. (4.20) in Ref. \onlinecite{risken1984fokker}) and $\dot{\vec{X}}\sim  \Delta \vec{X}/\Delta t$, Eq. \eqref{eq-32} is approximated by  
\begin{equation}\label{eq-42}
\frac{\partial f(\vec{x},t)}{\partial t}
=
-\nabla_{\vec{x}} \langle\dot{\vec{X}}(t)\rangle f(\vec{x},t)
+
\nabla_{\vec{x}}
\left(\mathscr{B}(\vec{x},t) 
\nabla_{\vec{x}}f(\vec{x},t)\right)
,
\end{equation}
where the drift coefficient is  
\begin{equation}\label{eq-43}
\langle\dot{\vec{X}}(t)\rangle= \lim_{\Delta t \to 0} \frac{\langle \Delta{\vec{X}}\rangle}{\Delta t},
\end{equation}
and the memory kernel is  
\begin{equation}\label{eq-44}
\mathscr{B}(\vec{x},t)=\frac{1}{2}\lim_{\Delta t \to 0} 
\frac
{\left\langle\Delta^2{\vec{X}} \right\rangle}{\Delta t},
\end{equation}
implying that when $t-s$ is infinitesimal, Eq. \eqref{eq-32} is also reduced to Markovian Fokker-Planck equation. 

Therefore, it is concluded that, in the white noise limit, the present study is reduced to the classical Fokker-Planck equation for Markovian processes. This conclusion is easy to understand, because the integral curve $\vec{S}$ can be represented by a small line segment $\Delta\vec{X}$ for an infinitesimal time interval $\Delta t$, and thus the cumulation with respect to the state transition paths is approximated by jump moments for Markovian processes.

%
%

\section{\label{sec:level4}Dispersion of Particles in Turbulent Flows}

%
%

Consider a special case of Eq. \eqref{eq-1}:
\begin{equation}\label{eq-45}
\frac{\mathrm{d} \vec{X} }{\mathrm{d}t}=   \vec{F}(\vec{X})+\boldsymbol{\xi}, 
\end{equation}
where $\vec{F}$ is assumed to be a linear function of $\vec{X}$, and $\boldsymbol{\xi}$ is a random forcing and unnecessarily Gaussian white noise as usually assumed.  Using Eq. \eqref{eq-25}, taking ensemble average on all possible paths leading to the point $\vec{x}$, i.e., $\vec{X}(\tau)=\vec{X}(\tau|\vec{X}(s),s)$ ($0 \le \tau \le t$, with boundary conditions of $\vec{x}=\vec{X}(t|\vec{X}(s),s)$, and $\vec{X}(s)=\vec{X}(s|\vec{X}(s),s)=0$ for simplicity), we obtained that the drift term is
\begin{eqnarray}\label{eq-46}
\langle \mathscr{D}^{(1)}(\vec{x},t)\rangle =\frac{\partial \left\langle \vec{S}(t) \right\rangle}{\partial t}   
=\vec{F}(\vec{x}) +\langle \boldsymbol{\xi}  \rangle.
\end{eqnarray}
If $\boldsymbol{\xi}$ is Gaussian, $\langle \boldsymbol{\xi}  \rangle=0$.
As to the diffusion term, its derivation is started from the expression for the path integration given by Eq. \eqref{eq-21}. It is
\begin{equation}\label{eq-47}
\vec{S}(t)= \int_{0}^t \dot{\vec{X}}(\tau)\mathrm{d}\tau = \langle \vec{S}(t)\rangle +\int_0^t \mathrm{d}\tau \vec{G}(t;\tau) \boldsymbol{\xi}(\tau),
\end{equation}
where the response function, or the Green function $\vec{G}(t;\tau)$, serving as a propagator to transfer impulses along particle paths, is determined by the following first-order ordinary differential equation:
\begin{equation}\label{eq-48}
\dot{\vec{G}}(t;\tau) = \vec{J} \cdot \vec{G}(t;\tau)+\vec{I}\delta(t-\tau),
\end{equation}
where $\vec{J} = \partial \vec{F}(\vec{X})/\partial \vec{X}|_{\vec{X}=\vec{x}}$.

Since that $\vec{G}(t;\tau)$ and $\langle \langle\boldsymbol{\xi}(\tau_1)\boldsymbol{\xi}(\tau_2)\rangle \rangle$ are functions of the time $t$, with the help of Leibniz integral rule, the diffusion coefficient is expanded in detail as follows: 
%
%

\begin{widetext}
\begin{eqnarray}\label{eq-49}
\langle \mathscr{D}^{(2)}(\vec{x},t)\rangle
&=&
\frac{1}{2!2!}\frac{\partial \left\langle \left\langle \vec{S}\vec{S} \right\rangle \right\rangle}{\partial t} \nonumber\\
&=&
\frac{1}{2!2!}\frac{\partial }{\partial t}
\int_0^t \int_{0}^t\mathrm{d}\tau_1\mathrm{d}\tau_2  \vec{G}(t;\tau_1) \langle \langle\boldsymbol{\xi}(\tau_1)\boldsymbol{\xi}(\tau_2)\rangle \rangle \vec{G}^T(t;\tau_2) \nonumber\\
&=&
+\frac{1}{2}  \int_0^t \mathrm{d}\tau  
\left(
\vec{G}(t;\tau) 
\langle \langle 
\boldsymbol{\xi}(\tau)\boldsymbol{\xi}(t)
\rangle \rangle
+ 
\langle \langle 
\boldsymbol{\xi}(t)\boldsymbol{\xi}(\tau)
\rangle \rangle \vec{G}^T(t;\tau)
\right)
\nonumber\\
&&+
\frac{1}{4} 
\int_0^t \int_{0}^t\mathrm{d}\tau_1\mathrm{d}\tau_2  
\vec{G}(t;\tau_1) \frac{\partial \langle \langle\boldsymbol{\xi}(\tau_1)\boldsymbol{\xi}(\tau_2)\rangle \rangle}{\partial t} \vec{G}^T(t;\tau_2)\nonumber\\ 
&& +
\frac{1}{4} \left(\vec{J}\langle \langle \vec{S}\vec{S} \rangle \rangle+\langle \langle \vec{S}\vec{S} \rangle \rangle\vec{J}^T \right) 
.
\end{eqnarray}
\end{widetext}
Eq. \eqref{eq-49} shows that the diffusion term involves the cumulant with respect to $\boldsymbol\xi$; therefore, it is unnecessary to assume that $\boldsymbol\xi$ is Gaussian. 

If a stationary turbulent flow is considered, so that $\langle \langle \boldsymbol{\xi}(\tau_1)\boldsymbol{\xi}(\tau_2)\rangle \rangle$ is independent of time $t$, then Eq. \eqref{eq-49} is reduced to 
\begin{widetext}
\begin{eqnarray}\label{eq-50}
\langle \mathscr{D}^{(2)}(\vec{x},t)\rangle 
&=&+
\frac{1}{2}  \int_0^t \mathrm{d}\tau  
\left(
\vec{G}(t;\tau) 
\langle \langle 
\boldsymbol{\xi}(\tau)\boldsymbol{\xi}(t)
\rangle \rangle
+ 
\langle \langle 
\boldsymbol{\xi}(t)\boldsymbol{\xi}(\tau)
\rangle \rangle \vec{G}^T(t;\tau)
\right)\nonumber\\
&&+
\frac{1}{4} \left(\vec{J}\langle \langle \vec{S}\vec{S} \rangle \rangle+\langle \langle \vec{S}\vec{S} \rangle \rangle\vec{J}^T \right) 
.
\end{eqnarray}
\end{widetext}
Furthermore, if the term of  $\vec{J}\langle \langle \vec{S}\vec{S} \rangle \rangle+\langle \langle \vec{S}\vec{S} \rangle \rangle\vec{J}^T$ is vanishingly small, Eq. \eqref{eq-49} is further reduced to 
\begin{widetext}
\begin{equation}\label{eq-51}
\langle \mathscr{D}^{(2)}(\vec{x},t)\rangle=
\frac{1}{2}  \int_0^t \mathrm{d}\tau  
\left(
\vec{G}(t;\tau) 
\langle \langle 
\boldsymbol{\xi}(\tau)\boldsymbol{\xi}(t)
\rangle \rangle
+ 
\langle \langle 
\boldsymbol{\xi}(t)\boldsymbol{\xi}(\tau)
\rangle \rangle \vec{G}^T(t;\tau)
\right)
.
\end{equation}     
\end{widetext}
Moreover, if $\vec{G}(t;\tau)$ is symmetric, i.e.,  $\vec{G}(t;\tau)=\vec{G}^T(t;\tau)$, denoting $\Delta \vec{X}(t)=\int_0^t \mathrm{d}\tau \vec{G}(t;\tau) \boldsymbol{\xi}(\tau)  $, Eq. \eqref{eq-51} has the form as follows
\begin{equation}\label{eq-52}
\langle \mathscr{D}^{(2)}(\vec{x},t)\rangle
=
\int_0^t \mathrm{d}\tau  
\vec{G}(t;\tau) 
\langle \langle 
\boldsymbol{\xi}(\tau)\boldsymbol{\xi}(t)
\rangle \rangle
=
\langle \langle 
\Delta \vec{X}(t) \boldsymbol{\xi}(t)
\rangle \rangle
,
\end{equation} 
which is the same as that has been derived by \citet{SWAILES199738} based on Furutsu–Novikov formula and  \citet{Reeks1991Kinetic,reeks1992continuume} based on LHDI method. 

The above analysis shows that, in comparison with previous studies, two new mechanisms contribute to diffusion in the phase space. The first one owes to the temporal variation of the autocorrelation of the random force $\boldsymbol{\xi}$. The second one arises from temporal variation of the response function $\vec{G}$, which indicates that changes in the propagator of turbulent impulses also leads to diffusion in the phase space.


Considering particles dispersion in a homogeneous turbulent field, one reads in Eq. \eqref{eq-45} that
\begin{widetext}
\begin{equation}\label{eq-53}
\vec{x}=
\left(                  
  \begin{array}{c}    
    \vec{r}\\  
    \vec{u}\\   
  \end{array}
\right) , 
\vec{X}=
\left(                  
  \begin{array}{c}    
    \vec{R}\\  
    \vec{U}\\   
  \end{array}
\right) , 
\boldsymbol{\xi}=
\left(                  
  \begin{array}{c}    
    \vec{0}\\  
    {\beta (\vec{V}-\langle \vec{V}\rangle )}\\   
  \end{array}
\right) , 
\vec{F}(\vec{X})=
\left(                  
  \begin{array}{c}    
    {\vec{U}}\\  
    -\beta ({\vec{U}}- \langle \vec{V} \rangle)\\   
  \end{array}
\right) , 
\end{equation}
\end{widetext}
where $\vec{V}=\vec{V}(\vec{X}(t))$ is the velocity of the carrier fluid seen by particles; $\vec{R}$ and $\vec{U}$ are respectively the position and velocity vector of parties; $\beta$ is is the inverse of the particle relaxation time.  
When it is assumed that $\beta$ is constant and the flows is homogeneous,  we found that the response function is given by Eq. \eqref{eq-48} as follows:
\begin{equation}\label{eq-54}
\vec{G}(t;\tau)=\beta^{-1}
\left(                  
  \begin{array}{cc}    
    0&\ 1-\mathrm{e}^{-\beta (t-\tau)} \\  
    0& \beta\mathrm{e}^{-\beta (t-\tau)}\\   
  \end{array}
\right) . 
\end{equation}


If further assumed that $\langle\langle \vec{V}(\vec{X}(t))\vec{V}(\vec{X}(t-\tau)) \rangle\rangle=\vec{D}\mathrm{e}^{-|\tau|/T_{Lp}}$ with  $T_{Lp}$ denoting the integral time scale of turbulence along the paths of particles, and $\vec{D}$ is assumed independent of time,
then we finally found that 
\begin{equation}\label{eq-55}
\langle \mathscr{D}^{(2)}(\vec{x},t)\rangle=   
\left(                  
  \begin{array}{cc}    
    \boldsymbol{\kappa} &\boldsymbol{\lambda}\\  
    \boldsymbol{\lambda} &\boldsymbol{\mu}\\   
  \end{array}
\right) 
, 
\end{equation}
where  
\begin{equation}\label{eq-56}
\boldsymbol\kappa =\frac{  \vec{D} T_{Lp} (1-\mathrm{e}^{-\beta t})}{ 2(1-\beta T_{Lp})}\left(1-\beta T_{Lp}+ \beta T_{Lp} \mathrm{e}^{-T_{Lp}^{-1} t}-\mathrm{e}^{-\beta t} \right),
\end{equation}
\begin{eqnarray}\label{eq-57}
\boldsymbol\lambda 
&=&
\frac{\beta \vec{D} T_{Lp} }{2}  \left[	1-\mathrm{e}^{-\frac{t}{T_{Lp}}}-\frac{1-\mathrm{e}^{-\left(\beta+T_{Lp}^{-1}\right)t}}{1+\beta T_{Lp}}\right] \nonumber\\
&+&\frac{\beta  \vec{D} T_{Lp}}{4(1-\beta T_{Lp})}\left[1-\mathrm{e}^{-2\beta t}-\frac{2\beta T_{Lp} \left(1-\mathrm{e}^{-(\beta+T_{Lp}^{-1})t}\right)}{1+\beta T_{Lp}}\right]\nonumber\\
&-&\frac{\beta \vec{D} T_{Lp}(1-\mathrm{e}^{-\beta t})}{4(1-\beta T_{Lp})}\left(1-\beta T_{Lp}+ \beta T_{Lp} \mathrm{e}^{-T_{Lp}^{-1} t}-\mathrm{e}^{-\beta t} \right),\nonumber\\
\end{eqnarray}
and 
\begin{eqnarray}\label{eq-58}
&&\boldsymbol\mu =\frac{\beta^2\vec{D}  T_{Lp}}{1+\beta T_{Lp}}\left[1-\mathrm{e}^{-\left(\beta+T_{Lp}^{-1}\right)t}\right]\nonumber\\
&-&\frac{1}{2}\frac{\beta^2 \vec{D} T_{Lp}}{1-\beta T_{Lp}}\left[1-\mathrm{e}^{-2\beta t}-\frac{2\beta T_{Lp} \left(1-\mathrm{e}^{-(\beta+T_{Lp}^{-1})t}\right)}{1+\beta T_{Lp}}\right]. \nonumber\\
\end{eqnarray}

When $t\to \infty$, the diffusion coefficients reach their stable values:  
\begin{equation}\label{eq-59}
\boldsymbol\kappa_{\infty} =\frac{ \vec{D} T_{Lp}  }{ 2 }, \boldsymbol\lambda_{\infty} 
= \frac{ \beta^2 \vec{D} T_{Lp}^2}{4(1+\beta T_{Lp})}, \boldsymbol\mu_{\infty} =\frac{\beta^2\vec{D}  T_{Lp}}{2(1+\beta T_{Lp})}.
\end{equation}

If we assumed that $\partial^2/\partial \vec{r}\partial \vec{u}=\partial^2/\partial \vec{u}\partial \vec{r}$, implying that $f(\vec{r},\vec{u},t)$ is continuous in the phase space, then the kinetic equation for the system determined by Eq. \eqref{eq-53} is given by
\begin{widetext}
\begin{equation}\label{eq-60}
\frac{\partial f(\vec{r},\vec{u},t)}{\partial t}
+
\frac{\partial \vec{u}f(\vec{r},\vec{u},t)}{\partial \vec{r}} 
+
 \frac{\partial \beta (\langle \vec{V} \rangle- \vec{u}) f(\vec{r},\vec{u},t)}{\partial \vec{u}} 
=
\left(
\boldsymbol{\kappa}\frac{\partial^2  }{\partial \vec{r}^2 }  
+
\boldsymbol{\gamma}\frac{\partial^2  }{\partial \vec{u}\partial \vec{r}}  
+
 \boldsymbol{\mu}\frac{\partial^2  }{\partial \vec{u}^2}
\right) 
f(\vec{r},\vec{u},t),
\end{equation}
\end{widetext}
where $\boldsymbol\gamma=2\boldsymbol{\lambda}$.
If we define that 
\begin{equation}\label{eq-61}
\alpha = \int{\mathrm{d}\vec{u}} f(\vec{r},\vec{u},t), \alpha \tilde{\vec{u} } =\int{\mathrm{d}\vec{u}} f(\vec{r},\vec{u},t) \vec{u} ,
\end{equation}
by virtue of Eq. \eqref{eq-60}, we had the equations of mass and momentum conservation for disperse phase as follows: 
\begin{equation}\label{eq-62}
\frac{\partial \alpha}{\partial t}+\frac{\partial \alpha  \tilde{\vec{u} }}{\partial \vec{r} } = \boldsymbol\kappa  \frac{\partial^2 \alpha}{\partial \vec{r}^2},
\end{equation}
and 
\begin{equation}\label{eq-63}
\frac{\partial \alpha \tilde{\vec{u} }}{\partial t}+\frac{\partial \alpha \widetilde{\vec{u}\vec{u} } }{\partial \vec{r}}  = 
\boldsymbol\kappa
\frac{\partial^2 \alpha \tilde{\vec{u} }}{\partial \vec{r}^2 }
- \boldsymbol\gamma \frac{\partial  \alpha }{\partial \vec{r}  }
+\alpha \beta (\langle \vec{V} \rangle-  \tilde{\vec{u} })
+\alpha \vec{b},
\end{equation}
respectively, where $\vec{b}$ is a body force on particles. It can be found that the diffusion additional to conventional research also affect macroscopic transport of particles as macroscopic  diffusions in both mass and momentum conservation equations. 


Dimensional analysis shows that the relative importance of the diffusion given by the terms on the right-hand-side of Eq. \eqref{eq-60} depends on the Stokes number $St=(\beta T_L)^{-1}$ and the parameter $\eta=T_{Lp}/T_L$, where $T_L$ is the Lagrangian integral time scale of fluid. The parameter $\eta=T_{Lp}/T_L$ is also a complicated function of the Stokes number and the particle-to-fluid density ratio\cite{OESTERLE2006838}. It shows that, for low inertia particles, i.e., the Stokes number $St \to 0$, $\eta \to O(1)$ \cite{OESTERLE2006838}, and thus the orders of magnitude of 
the terms on the right-hand-side of Eq. \eqref{eq-60} are $O(1)$, $O(St^{-1})$, and $O(St^{-1})$, respectively. While for the  high  inertia particles, when $St \to \infty$, the orders of magnitude are $O(\eta)$, $O(St^{-2}\eta^2)$, and $O(St^{-2}\eta)$, respectively, depending on both the Stokes number and the particle-to-fluid density ratio. For a special case of $\eta=1$, implying that particles are in neutral buoyancy\cite{OESTERLE2006838}, it shows that the first term is important for high inertia particles.  These results implies at least that the first term on the right-hand-side of Eq. \eqref{eq-60} is insignificant for low inertia particles.  

\begin{figure*}[h]
   \centering
   \subfloat[$St=0.1$]
   {
   	\label{fig:Fig2:a}
   	\begin{minipage}[c]{0.333333\textwidth}
   		\centering
   		\includegraphics[width=\textwidth]{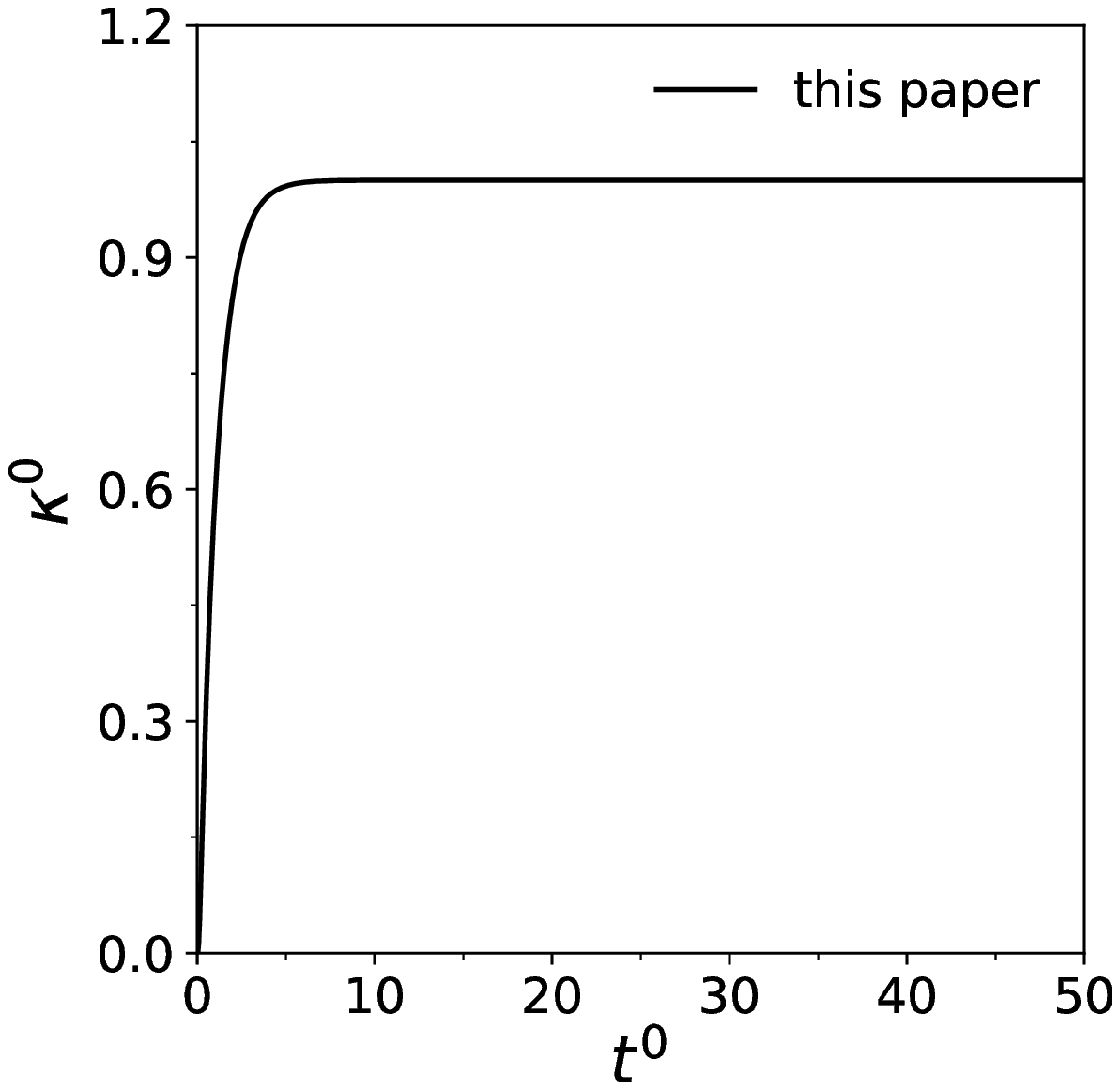}
   	\end{minipage}
   }   
   \subfloat[$St=1.0$]
   {
   	\label{fig:Fig2:b}
   	\begin{minipage}[c]{0.333333\textwidth}
  		 \centering
   		\includegraphics[width=\textwidth]{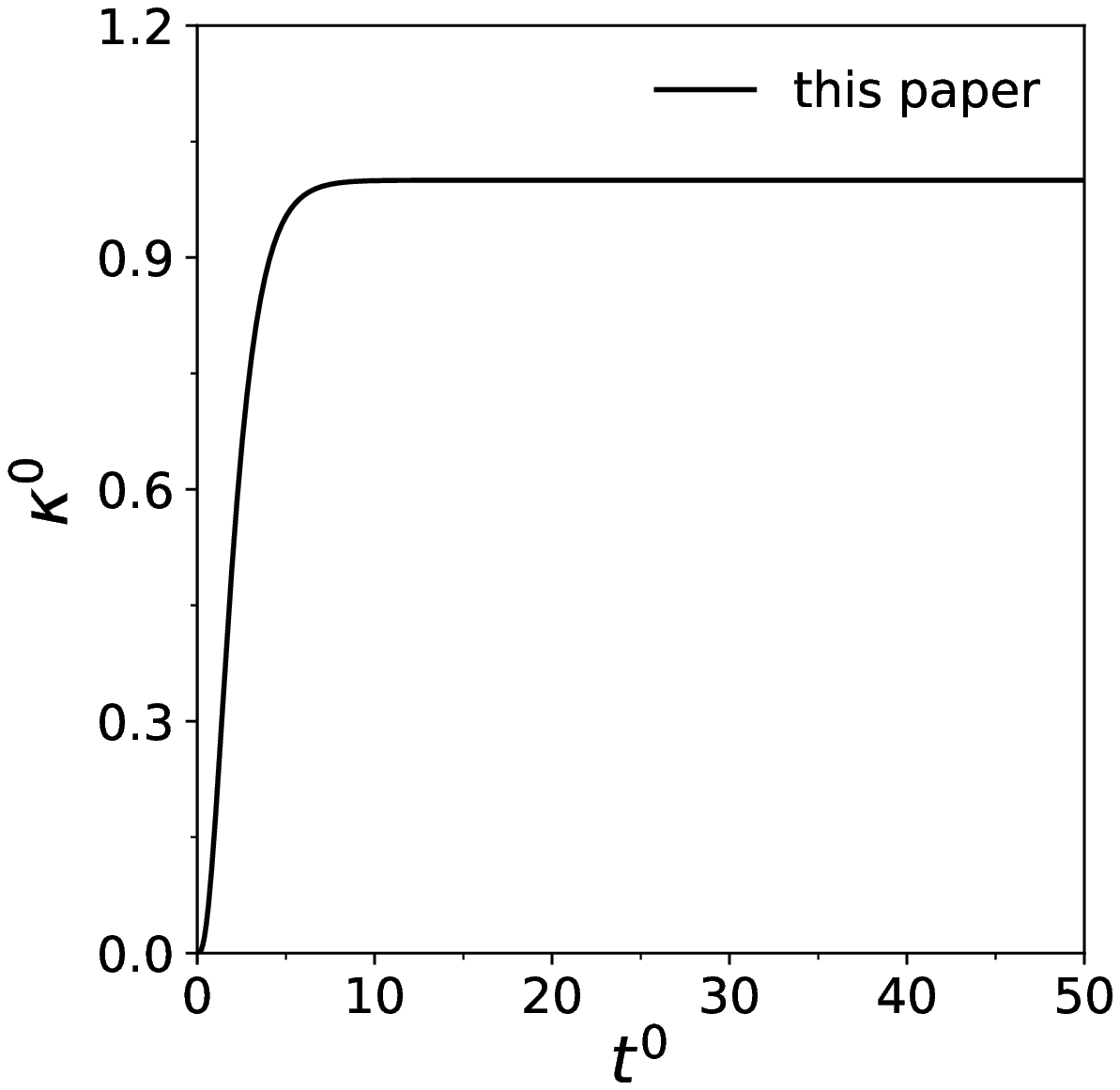}
   	\end{minipage}
   }     
   \subfloat[$St=10.0$]
   {
   	\label{fig:Fig2:c}
   	\begin{minipage}[c]{0.333333\textwidth}
   		\centering
   		\includegraphics[width=\textwidth]{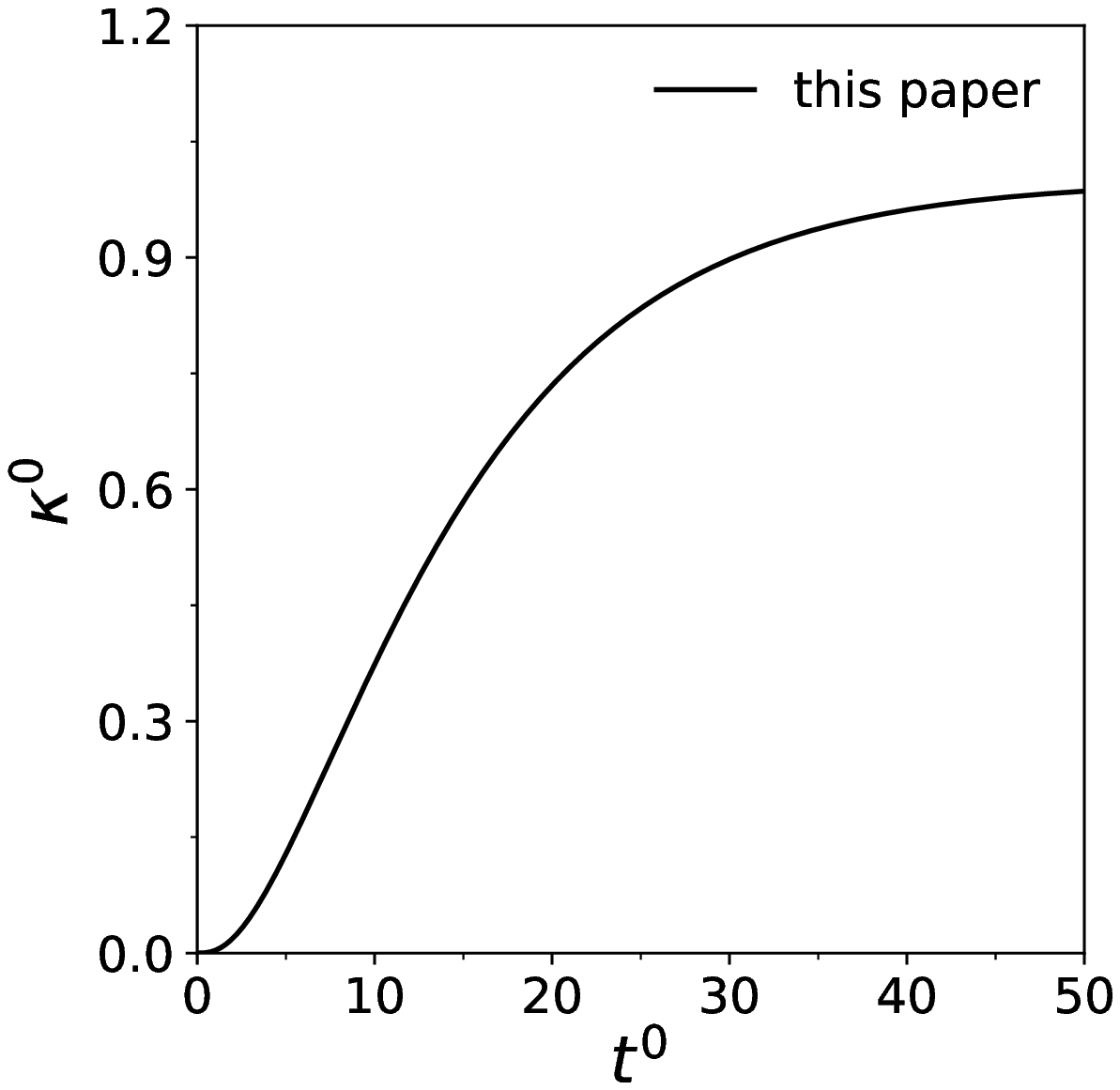}
   	\end{minipage}
   } 
   \\
   \subfloat[$St=0.1$]
   {
   	\label{fig:Fig2:d}
   	\begin{minipage}[c]{0.333333\textwidth}
   		\centering
   		\includegraphics[width=\textwidth]{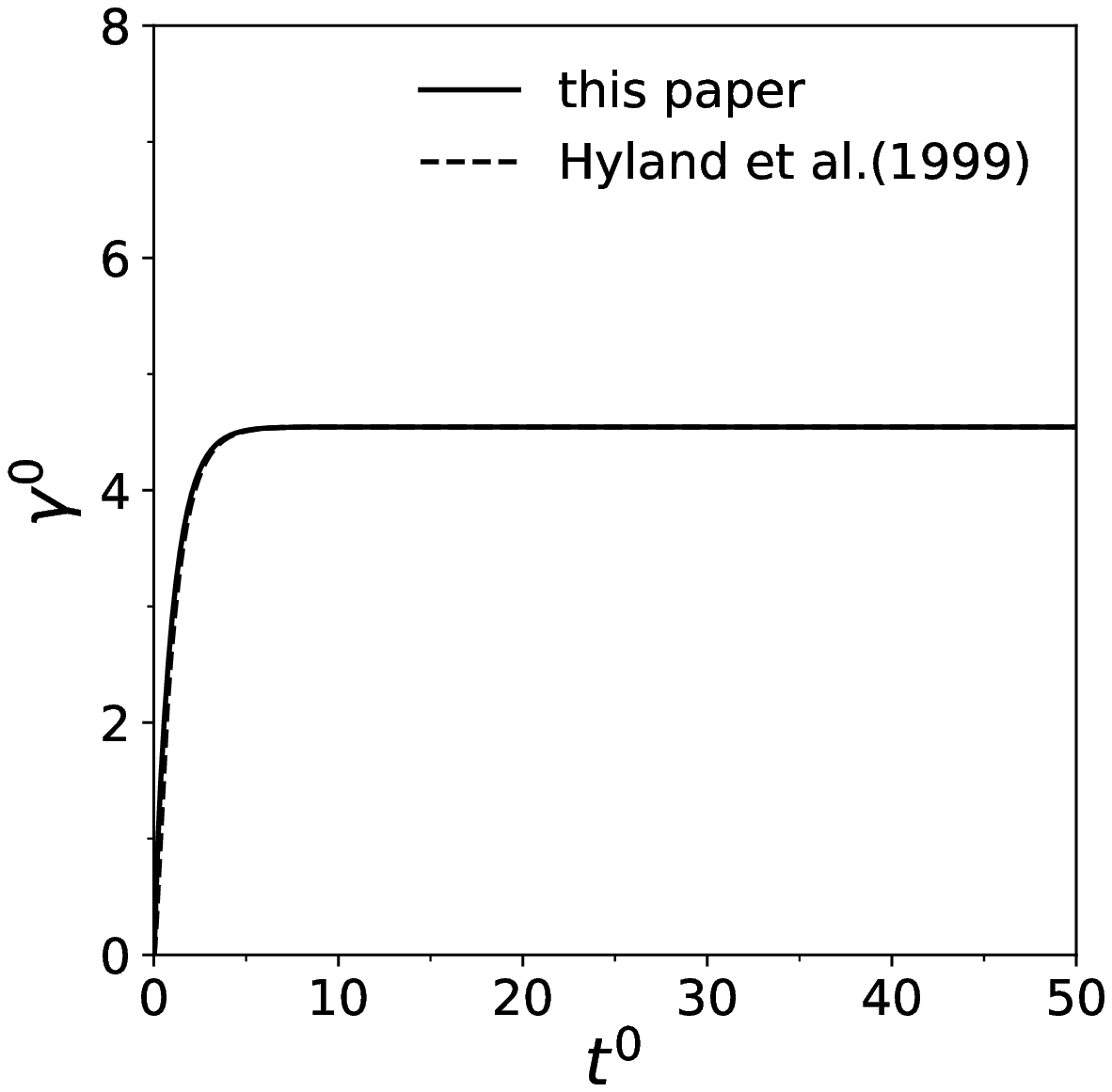}
   	\end{minipage}
   }   
   \subfloat[$St=1.0$]
   {
   	\label{fig:Fig2:e}
   	\begin{minipage}[c]{0.333333\textwidth}
  		 \centering
   		\includegraphics[width=\textwidth]{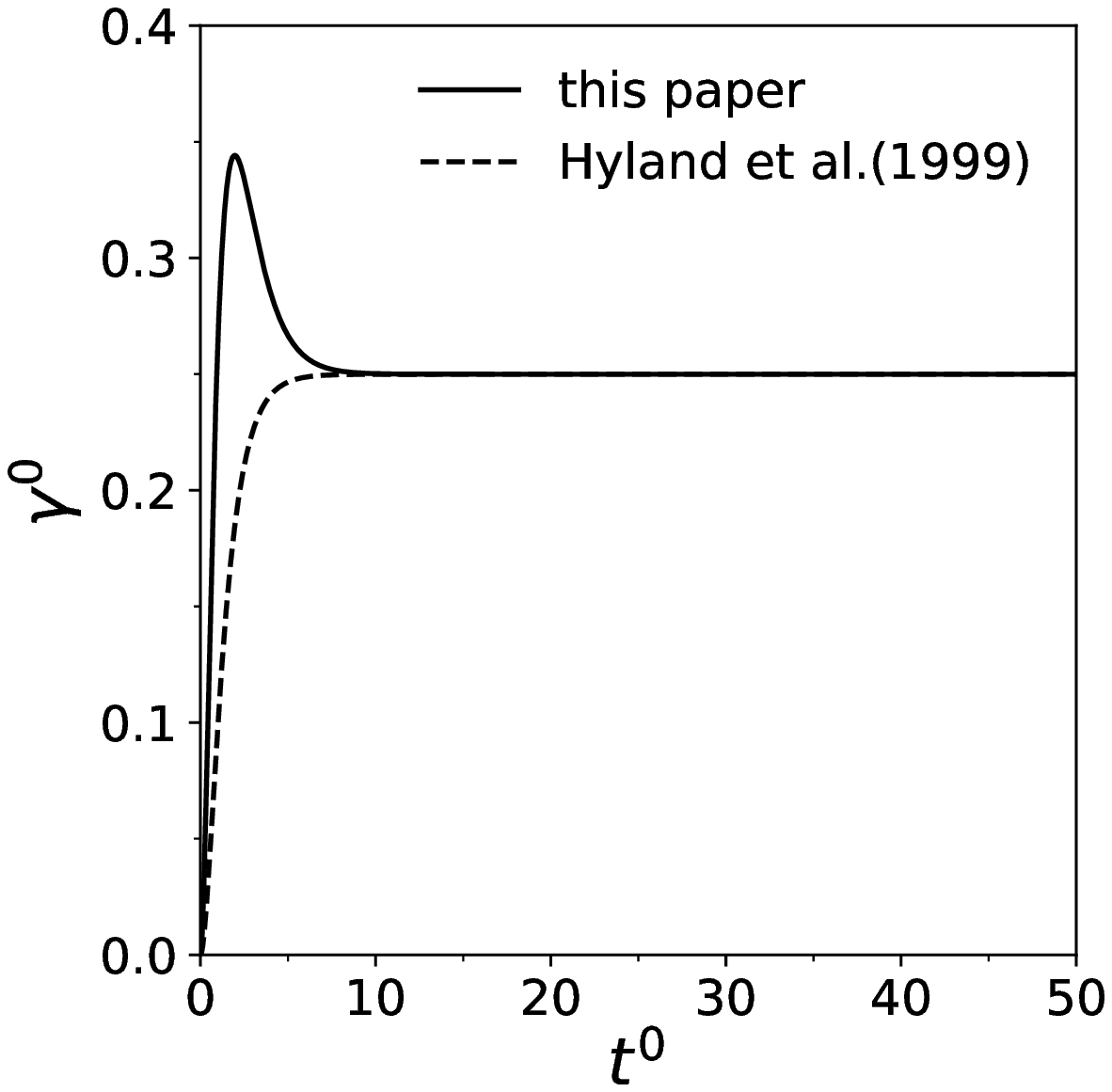}
   	\end{minipage}
   }     
   \subfloat[$St=10.0$]
   {
   	\label{fig:Fig2:f}
   	\begin{minipage}[c]{0.333333\textwidth}
   		\centering
   		\includegraphics[width=\textwidth]{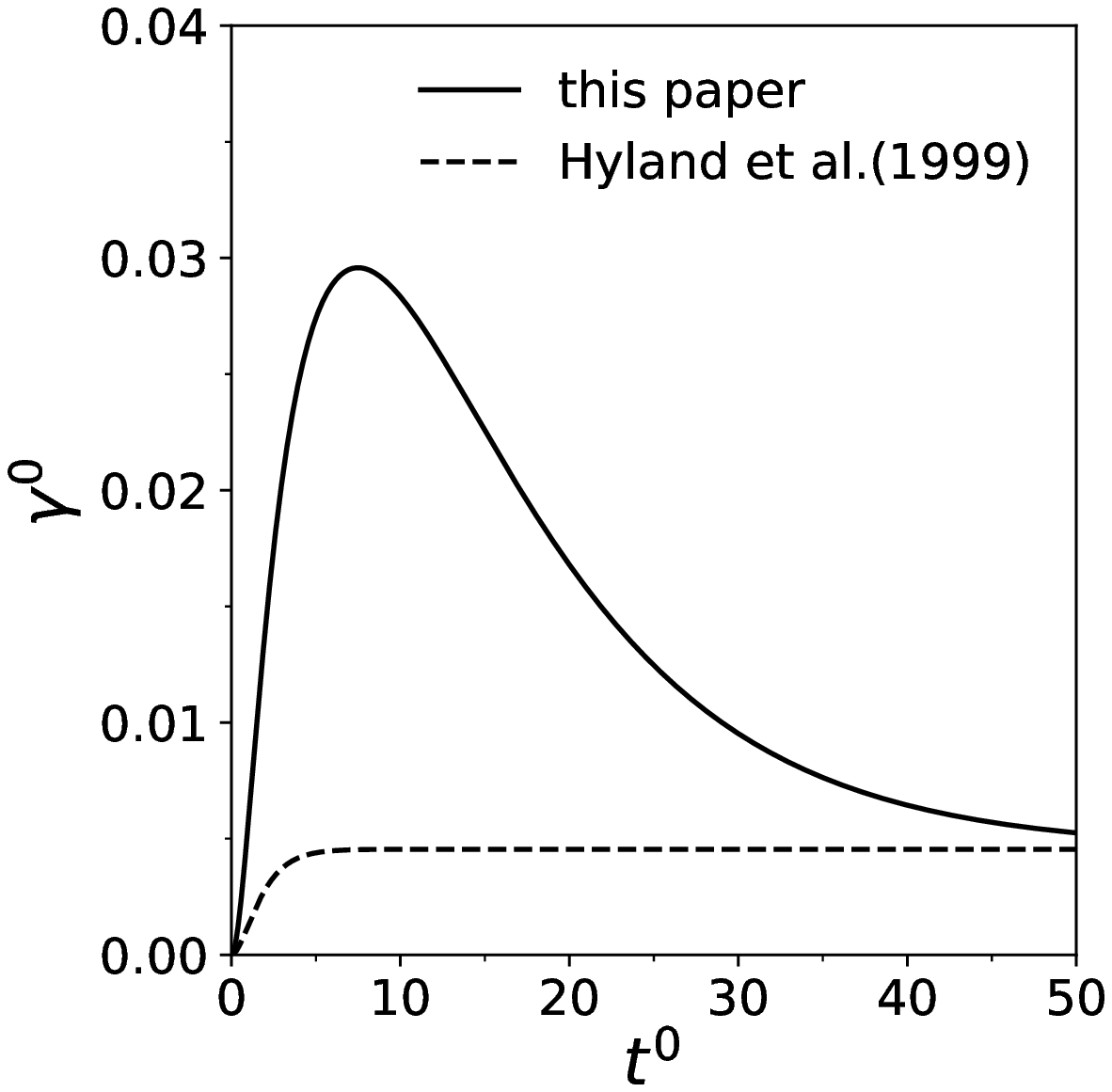}
   	\end{minipage}
   } 
   \\
   \subfloat[$St=0.1$]
   {
   	\label{fig:Fig2:g}
   	\begin{minipage}[c]{0.333333\textwidth}
   		\centering
   		\includegraphics[width=\textwidth]{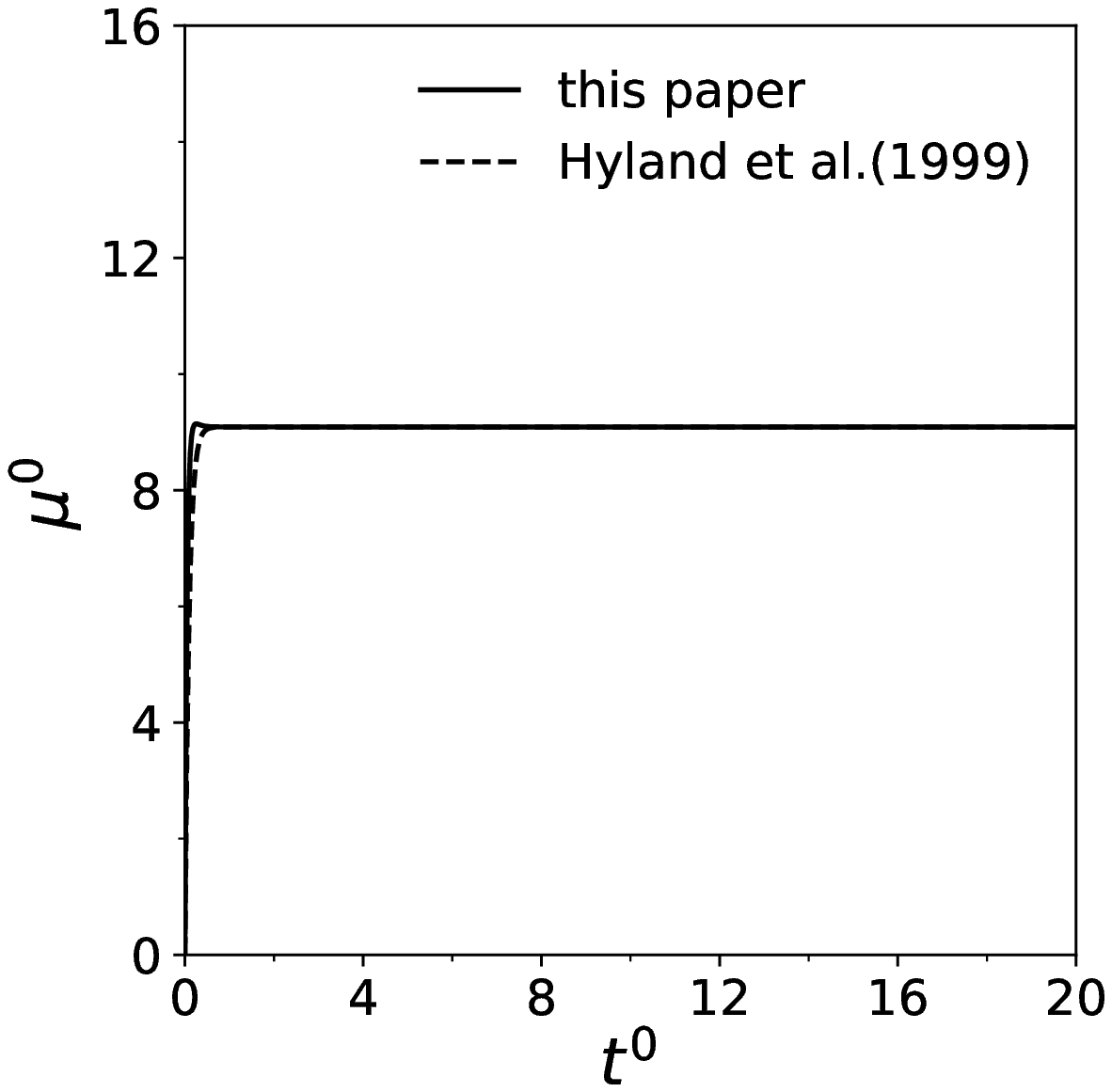}
   	\end{minipage}
   }   
   \subfloat[$St=1.0$]
   {
   	\label{fig:Fig2:h}
   	\begin{minipage}[c]{0.333333\textwidth}
  		 \centering
   		\includegraphics[width=\textwidth]{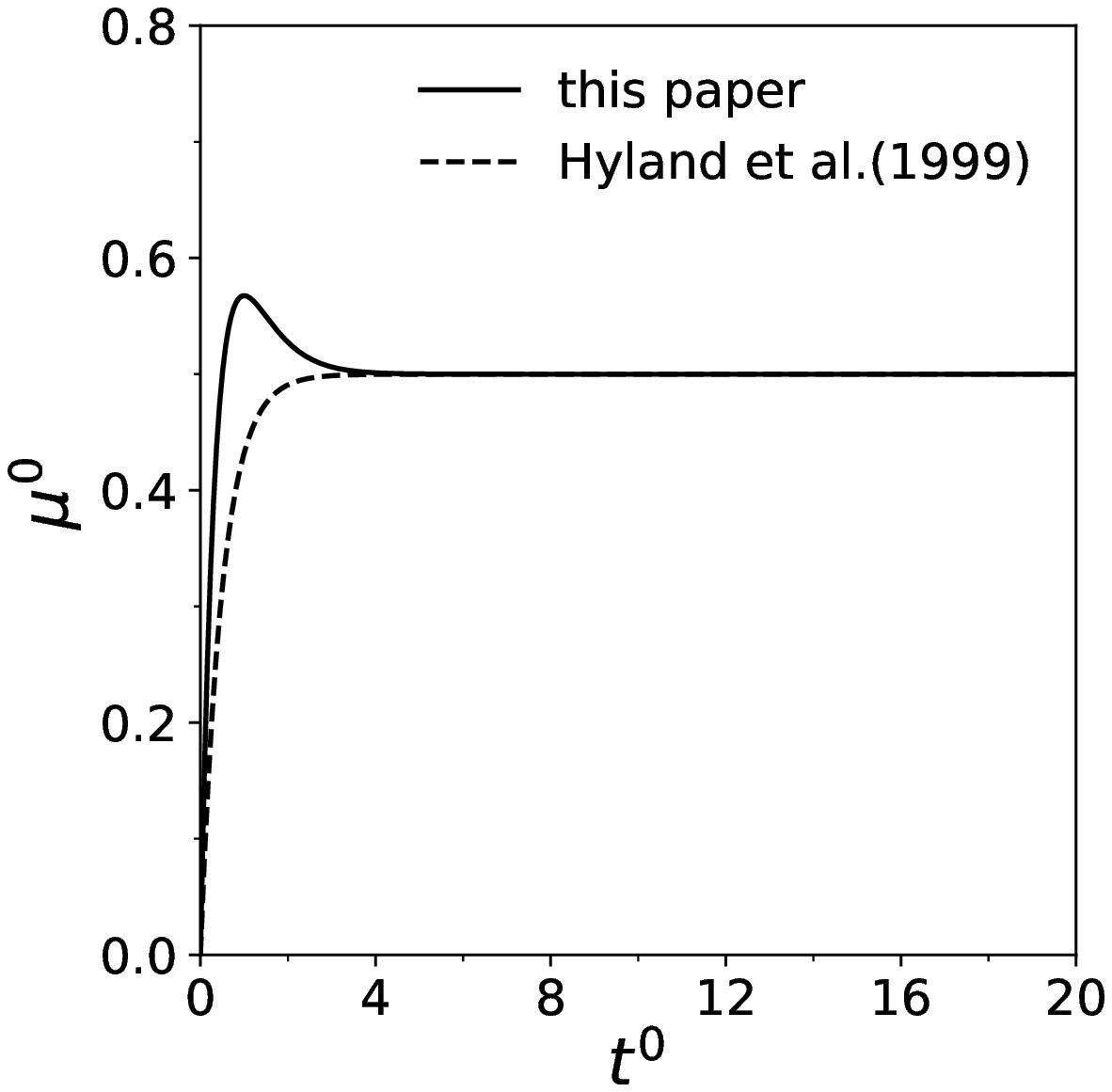}
   	\end{minipage}
   }     
   \subfloat[$St=10.0$]
   {
   	\label{fig:Fig2:i}
   	\begin{minipage}[c]{0.333333\textwidth}
   		\centering
   		\includegraphics[width=\textwidth]{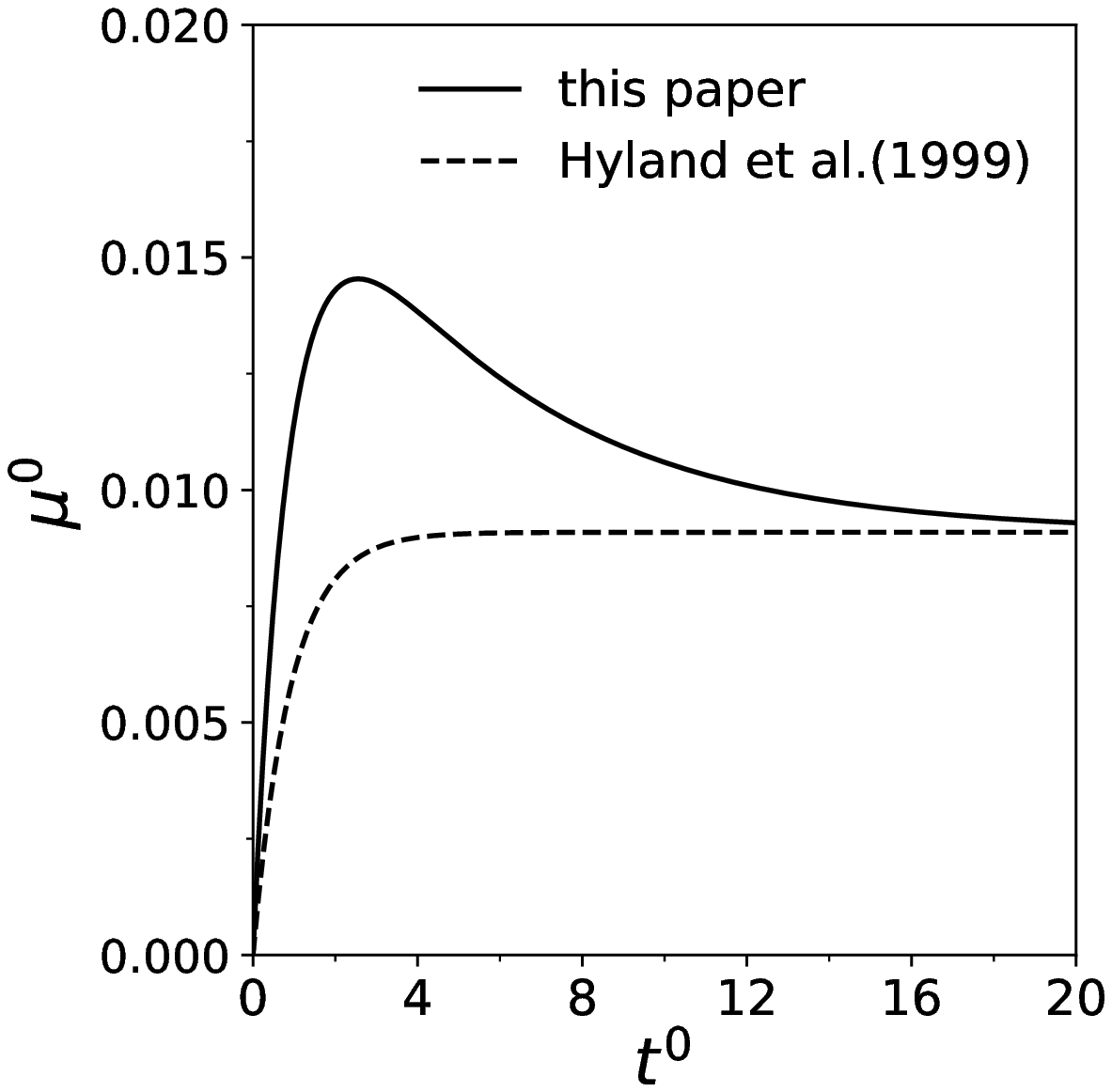}
   	\end{minipage}
   } 
\caption{Variation of the dimensionless diffusion coefficients $\boldsymbol\kappa^0=\boldsymbol\kappa\boldsymbol\sigma^{-1}T_L^{-1}$, $\boldsymbol\gamma^0=\boldsymbol\gamma\boldsymbol\sigma^{-1}$, and $\boldsymbol\mu^0=\boldsymbol\mu\boldsymbol\sigma^{-1}T_L$ derived in this paper against $t^0= T_L^{-1}t$ for different Stokes number $St=(\beta T_L)^{-1}$. Particles are assumed in neutral buoyancy. Solid lines are derived in this paper, and the dashed lines are those obtained in Ref. \onlinecite{hyland1999exact}}
\label{fig:Fig2} 
\end{figure*}

Fig. \ref{fig:Fig2} depicts  the variations of the dimensionless diffusion coefficients $\boldsymbol\kappa^0=\boldsymbol\kappa\boldsymbol\sigma^{-1}T_L^{-1}$, $\boldsymbol\gamma^0=\boldsymbol\gamma\boldsymbol\sigma^{-1}$, and $\boldsymbol\mu^0=\boldsymbol\mu\boldsymbol\sigma^{-1}T_L$  against $t^0= T_L^{-1}t$ for different Stokes number $St$, with $\boldsymbol\sigma=\vec{D}/2$. For the purpose of simplicity, it is assumed $\eta =1$, which is true for particles in  neutral buoyancy\cite{OESTERLE2006838}.  It shows that for the case of $St=0.1$, $\boldsymbol\kappa^{0}$ is  one oder of magnitude lower than those of $\boldsymbol\gamma^{0}$ and $\boldsymbol\mu^{0}$; differently, for the situation of $St=10.0$, $\boldsymbol\kappa^{0}$ is  two orders of magnitude higher than that of $\boldsymbol\gamma^{0}$ and $\boldsymbol\mu^{0}$. This fact implies that the first term on the right-hand-side of Eq. \eqref{eq-60} is important for high inertial parties, but has a neglectable effect on low inertia particles. In addition, the results obtained in Ref. \onlinecite{hyland1999exact} are also plotted in Fig.2. It shows that for $St=0.1$, $\boldsymbol\gamma^{0}$ and $\boldsymbol\mu^{0}$ given by the present study has nearly the same variation as that of Ref. \onlinecite{hyland1999exact}. While for high inertia case, e.g., $St=10.0$, both $\boldsymbol\gamma^{0}$ and $\boldsymbol\mu^{0}$ increase sharply to reach their maximums and then decrease slowly to their stable values, which were not observed in previous studies. 
%
%


%
%
%
\section{\label{sec:level5}Concluding remarks}

PDF formulation of disperse two-phase turbulent flows has attracted considerable attention in the past decades. The key difficulty encountered is how to close a turbulent diffusion term in the phase space arising from ensemble average on the conservation equation for fine-grained phase-space density function.  This difficulty has been extensively investigated in the past, and different theories and methods are established to cope with this difficulty, while it nevertheless remains as an open question for further study.

This study aimed to derive a kinetic equation for particle dispersion in turbulent flows with a new approach. A local path density operator is introduced to identify state transition paths, and based on which the probability density function is expanded as a series in terms of the cumulants with respect to particle paths in the phase space. With this expansion, a kinetic equation with the diffusion terms in closed forms is directly obtained. This study shows that the derived kinetic equation is similar in form to the previous studies, but with its coefficients expressed in terms of the cumulants of particle paths in the phase space. It also shows that the present study possesses the features of:
\begin{enumerate}
\item The turbulent diffusion terms are derived directly in closed form by a series expansion of the probability density function in terms of the cumulants of particle paths;  
\item The kinetic equation is applicable to non-Markovian processes; whilst in the white noise limit, it is reduced to the classical Fokker-Planck equation; 
\item There are two new mechanisms that contribute to diffusion in the phase space which have not been reported in the past. 
\end{enumerate}

This study is solely focused on deriving a kinetic equation for disperse of particles in turbulent flows by a new approach. For simplicity, phenomena such as particle-particle interactions, thermal forcing, and phasic change are not included in the present formulation. In addition, a simple application to particle-laden flows is presented to examine its fundamental features in simple flow conditions, while further discussions on general features of disperse two-phase flows have not been touched. More comprehensive studies by means of the kinetic equation derived in this study will be our future tasks.

\begin{acknowledgments}
	This study is supported by National Natural Science Foundation of China with Grant No. 91547204.
\end{acknowledgments}

\appendix

\section{\label{sec:A-A}Proof of Eq. \eqref{eq-30}}
Since that 
\begin{eqnarray}\label{eq-A1}
\nabla_{\vec{x}}f(\vec{x},t|\vec{y},s)
&=&\nabla_{\vec{x}} \langle \chi(|\vec{x}-\vec{X}(t|\vec{y},s)|) \rangle\nonumber\\
&=& \langle \mathscr{U}(t|s) \rangle \nabla_{\vec{x}}\chi(|\vec{x}-\vec{y}|),
\end{eqnarray}
Using the relation $\nabla_{\vec{x}}\chi(|\vec{x}-\vec{y}|)=- \nabla_{\vec{y}}\chi(|\vec{x}-\vec{y}|)$, we found that
\begin{eqnarray}\label{eq-A2}
\nabla_{\vec{x}}f(\vec{x},t|\vec{y},s)
&=&- \nabla_{\vec{y}} \langle \mathscr{U}(t|s) \rangle\chi(|\vec{x}-\vec{y}|)\nonumber\\
&=&- \nabla_{\vec{y}}f(\vec{x},t|\vec{y},s).
\end{eqnarray}

\section{\label{sec:A-B}Proof of Eq. \eqref{eq-31}}
Since that 
\begin{eqnarray}\label{eq-B1}
&&(-1)^n \nabla_{\vec{x}}^n  \left\langle \mathscr{D}^{(n)}(\vec{x},t|\vec{y},s)\right\rangle  f(\vec{x},t|\vec{y},s)\nonumber\\
&=&(-1)^n \sum_{k=0}^n C_n^k \nabla_{\vec{x}}^{n-k} \left\langle \mathscr{D}^{(n)}(\vec{x},t|\vec{y},s)\right\rangle \nabla_{\vec{x}}^{k}f(\vec{x},t|\vec{y},s),
\end{eqnarray}
at the same time, with the help of Eq. \eqref{eq-A2}
\begin{eqnarray}\label{eq-B2}
&&\nabla_{\vec{y}}^n  \left\langle \mathscr{D}^{(n)}(\vec{x},t|\vec{y},s)\right\rangle  f(\vec{x},t|\vec{y},s)\nonumber\\
&=& \sum_{k=0}^n (-1)^k C_n^k \nabla_{\vec{y}}^{n-k} \left\langle \mathscr{D}^{(n)}(\vec{x},t|\vec{y},s)\right\rangle \nabla_{\vec{x}}^{k}f(\vec{x},t|\vec{y},s),
\end{eqnarray}
Using Eq. \eqref{eq-27} in conjunction with Eqs. \eqref{eq-B1} and \eqref{eq-B2}, we had Eq. \eqref{eq-31}.

\section{\label{sec:A-C}Derivation of Eq. \eqref{eq-32}}
 
Because $s$ in Eq. \eqref{eq-29} is arbitrary, let $s=t-\Delta \tau_L$, the following integration is approximated by:
\begin{eqnarray}\label{eq-C1}
	&&\int_s^t \mathrm{d}\tau \langle \langle \dot{\vec{X}}(\tau|\vec{y},s)\dot{\vec{X}}(t|\vec{y},s) \rangle \rangle\nonumber\\
	&=&
	\int_{t-\Delta \tau_L}^t \mathrm{d}\tau \langle \langle \dot{\vec{X}}(\tau|\vec{y},s)\dot{\vec{X}}(t|\vec{y},s) \rangle \rangle\nonumber\\
	&\approx&
	\Delta \tau_L \langle \langle \dot{\vec{X}}(t-\Delta \tau_L|\vec{y},t-\Delta \tau_L)\dot{\vec{X}}(t|\vec{y},t-\Delta \tau_L) \rangle \rangle.
\end{eqnarray}
Moreover, we assumed that $\langle \langle \dot{\vec{X}}(t-n\Delta \tau_L|\vec{y},t-n\Delta \tau_L)\dot{\vec{X}}(t|\vec{y},t-n\Delta \tau_L) \rangle \rangle \sim e^{-n \Delta \tau_L}$, and if $\Delta \tau_L$ satisfies that
$\langle \langle \dot{\vec{X}}(t-n\Delta \tau_L|\vec{y},t-n\Delta \tau_L)\dot{\vec{X}}(t|\vec{y},t-n\Delta \tau_L) \rangle \rangle\to 0$ for $n\ge 2$, then adding  $\Delta \tau_L\sum_{n=2}^N\langle \langle \dot{\vec{X}}(t-n\Delta \tau_L|\vec{y},t-n\Delta \tau_L)\dot{\vec{X}}(t|\vec{y},t-n\Delta \tau_L) \rangle \rangle$ to the left of the above equation does not change its value, so that
\begin{eqnarray}\label{eq-C2}
	&&\int_s^t \mathrm{d}\tau \langle \langle \dot{\vec{X}}(\tau|\vec{y},s)\dot{\vec{X}}(t|\vec{y},s) \rangle \rangle\nonumber\\
	&=&
	\int_{t-\Delta \tau_L}^t \mathrm{d}\tau \langle \langle \dot{\vec{X}}(\tau|\vec{y},s)\dot{\vec{X}}(t|\vec{y},s) \rangle \rangle\nonumber\\
	&\approx&
	\Delta \tau_L \langle \langle \dot{\vec{X}}(t-\Delta \tau_L|\vec{y},t-\Delta \tau_L)\dot{\vec{X}}(t|\vec{y},t-\Delta \tau_L) \rangle \rangle\nonumber\\
	&+&
	\Delta \tau_L \langle \langle \dot{\vec{X}}(t-2\Delta \tau_L|\vec{y},t-2\Delta \tau_L)\dot{\vec{X}}(t|\vec{y},t-2\Delta \tau_L) \rangle \rangle\nonumber\\
	&+&
	\cdots\nonumber\\
	&+&
	\Delta \tau_L \langle \langle \dot{\vec{X}}(t-N\Delta \tau_L|\vec{y},t-N\Delta \tau_L)\dot{\vec{X}}(t|\vec{y},t-N\Delta \tau_L) \rangle \rangle\nonumber\\
	&=&
	\sum_{n=1}^N
	\Delta \tau_L \langle \langle \dot{\vec{X}}(t-n\Delta \tau_L|\vec{y},t-n\Delta \tau_L)\dot{\vec{X}}(t|\vec{y},t-n\Delta \tau_L) \rangle \rangle.\nonumber\\
\end{eqnarray}
Let $\Delta \tau_L= t/N$, when $N\to \infty$, we found that 
\begin{eqnarray}\label{eq-C3}
	&&\lim_{N\to\infty}\sum_{n=1}^N
	\Delta \tau_L \langle \langle \dot{\vec{X}}(t-n\Delta \tau_L|\vec{y},t-n\Delta \tau_L)\dot{\vec{X}}(t|\vec{y},t-n\Delta \tau_L) \rangle \rangle\nonumber\\
	&=&
	\int_{0}^{t} \mathrm{d}\tau_L 
	\langle \langle\dot{\vec{X}}(t-\tau_L|\vec{y},t- \tau_L)\dot{\vec{X}}(t|\vec{y},t-\tau_L) \rangle \rangle
	\nonumber\\
        &=&
	\int_0^t \mathrm{d}\tau \langle \langle \dot{\vec{X}}(\tau|\vec{y},\tau)\dot{\vec{X}}(t|\vec{y},\tau) \rangle \rangle.
\end{eqnarray}
Using Eq. \eqref{eq-C2} and \eqref{eq-C3}, we had that   
\begin{equation}\label{eq-C4}
\int_s^t \mathrm{d}\tau \langle \langle \dot{\vec{X}}(\tau|\vec{y},s)\dot{\vec{X}}(t|\vec{y},s) \rangle \rangle
	 = 
\int_0^t \mathrm{d}\tau \langle \langle \dot{\vec{X}}(\tau|\vec{y},\tau)\dot{\vec{X}}(t|\vec{y},\tau) \rangle \rangle.
\end{equation}
Substitution of Eq. \eqref{eq-C4} into Eq. \eqref{eq-29} leads to Eq. \eqref{eq-32}. 
\bibliography{bib2019.bib}

\end{document}